\documentclass[preprint,12pt]{elsarticle}

\usepackage{graphicx}
\usepackage{amssymb}

\usepackage{graphicx}
\usepackage{dcolumn}
\usepackage{bm}
\usepackage{braket}
\usepackage{graphicx}
\usepackage{epstopdf}
\usepackage{mwe}    
\usepackage{subfig}
\usepackage[T1]{fontenc}
\usepackage{caption}
\usepackage{amssymb}
\usepackage{amsmath}
\usepackage{booktabs}
\usepackage{tabularx}
\usepackage{multirow}
\usepackage{caption}
\usepackage{floatrow}
\usepackage{epstopdf}
\usepackage{xcolor}

\newcommand{\tensor}[1]{\bm{\mathsf{#1}}} 

\journal{Journal Name}

\begin{document}

\begin{frontmatter}


\title{Cascaded Lattice Boltzmann Modeling and Simulations of Three-Dimensional Non-Newtonian Fluid Flows}


\author{Saad  Adam}
\ead{saad.adam@ucdenver.edu}
\author{Farzaneh  Hajabdollahi}
\ead{Farzaneh.Hajabdollahi-Ouderji@ucdenver.edu}
\author{Kannan N. Premnath}
\ead{kannan.premnath@ucdenver.edu}

\address{Department of Mechanical Engineering, University of Colorado Denver, 1200 Larimer street, Denver, Colorado 80217 , U.S.A}



\begin{abstract}
Non-Newtonian fluid flows, especially in three dimensions (3D), arise in numerous settings of interest to physics. Prior studies using the lattice Boltzmann method (LBM) of such flows have so far been limited to mainly to two dimensions and used less robust collision models. In this paper, we develop a new 3D cascaded LBM based on central moments and multiple relaxation times (MRT) on a three-dimensional, nineteen velocity (D3Q19) lattice for simulation of generalized Newtonian (power law) fluid flows. The relaxation times of the second order moments are varied locally based on the local shear rate and parameterized by the consistency coefficient and the power law index of the nonlinear constitutive relation of the power law fluid. Numerical validation study of the 3D cascaded LBM for various benchmark problems, including the complex 3D non-Newtonian flow in a cubic cavity at different Reynolds numbers and power law index magnitudes encompassing shear thinning and shear thickening fluids, are presented. Furthermore, in order to demonstrate the advantages of the proposed 3D cascaded LBM based on central moments, numerical stability comparisons against the LBMs based on a single relaxation time model and a MRT model using raw moments are made. Numerical results demonstrate the accuracy, second order grid convergence and significant improvements in numerical stability of the 3D cascaded LBM for simulation of 3D non-Newtonian flows of power law fluids.

\end{abstract}




\end{frontmatter}



\section{Introduction}

In the last few decades, the lattice Boltzmann Method (LBM) has become a preferred method for simulating complicated physical, chemical, and fluid mechanics problems~\cite{benzi1992lattice,chen1998lattice,succi2001lattice,aidun2010,kruger2017lattice}. It is a kinetic-based approach for fluid flow computations. This is why it is especially useful for computing fluid flows with multiple components involving interfacial dynamics, nonlinear constitutive models and complex boundaries. In particular, see the earliest review~\cite{benzi1992lattice}, which remains a source of original ideas and has contributed towards number of advances in this field. Because the LBM lies on the scale of mesoscopic level, challenges that have been encountered with using conventional CFD methods are not found with the LBM. The lattice Boltzmann Equation (LBE) can be constructed through several ways. A pioneering top-down formulation of the LBE with the desired macroscopic behavior for efficient flow simulations was presented in~\cite{higuera1989lattice}. Alternatively, one can derive the LBE through a dramatically simplified version of the Boltzmann kinetic equation. In addition, there are several approaches to perform mathematical analysis and derive the Navier-Stokes equations from the LBE. With Chapman-Enskog expansion being more popular among the rest, other approaches include the asymptotic expansion, extended Taylor series expansion and order of magnitude analysis.

The lattice Boltzmann methods are comprised of two fundamental steps, which are the streaming step and collision step. The streaming step is the same in various models of the LBM. However, because the collision step is more complicated, researchers have been devoting considerable efforts into finding the most suitable collision model for the lattice Boltzmann method. Among those different collision models, the simplest and commonly  used is the so-called single-relaxation-time (SRT) model~\cite{qian1992lattice}, which is based on the Bhatnagar-Gross-Krook (BGK) approximation~\cite{bhatnagar1954model}. On the other hand, a multiple-relaxation-time (MRT) model has been developed in order to improve numerical stability~\cite{d2002multiple}. MRT has been confirmed as a more stable collision model in various problems. In the MRT collision model, various moments are relaxed to their equilibrium states at different rates during the collision step. Alternatively, nonlinear stability can be achieved by the H-theorem compliant entropic lattice
Boltzmann methods (see e.g.,~\cite{karlin1999perfect,chikatamarla2006entropic,bosch2015entropic,frapolli2016entropic}). The equilibrium distribution of
the particle populations is constructed to minimize a convex entropy function by subjecting it to the constraints of the local conservation laws of mass and momentum. The use of entropy estimates to construct the post-collision states provides one of the most efficient LB methods for simulating low viscous flows.

More recently, a new class of collision model, the so-called cascaded LBM, has been introduced by Geier et al.~\cite{Geier2006}. Later on, Asinari~\cite{Asinari2008} reinterpreted this approach based on  relaxation to a generalized equilibrium. Premnath and Banerjee~\cite{premnath2009incorporating} incorporated forcing terms in the cascaded LBM by the method of central moments and systematically derived it and demonstrated its consistency to the Navier-Stokes equations via a Chapman-Enskog expansion. In such an approach, Galilean invariance is naturally enforced to the lattice Boltzmann equation (LBE) based on the relaxation of central moments. This involves computing moments which are shifted by the macroscopic fluid velocity. In another words, the moments are prescribed in a moving frame of reference. Comparatively, the moments in the prior approaches are computed in a rest frame of reference, which are termed as the raw moments.

In various recent studies~\cite{Geier2015,Ning2015}, the cascaded LBM based on central moments and multiple relaxation times has been shown to be significantly more stable when compared to the SRT collision model based LBM for the simulation of Newtonian fluid flows. Three-dimensional central moment LBMs including the forcing terms for D3Q15 and D3Q27 lattices were systematically derived in~\cite{premnath2011three}. A variant of such a formulation involving discrete moment equilibria rather than the continuous Maxwellian equilibria is presented in~\cite{Rosis16}.  A preconditioned cascaded LBM to accelerate steady flow simulations with   improved accuracy was recently developed by~\cite{hajabdollahi2018galilean}. Furthermore, the cascaded LBM has recently been extended to simulate thermal convective flows in two-dimensions (2D)~\cite{fei2018modeling,elseid2018cascaded}  and three-dimensions (3D) ~\cite{HAJABDOLLAHI2018838}. In addition, a modified forcing formulation and implementation of the central moment LBM has been presented in~\cite{fei2018three}. A systematic survey of various forcing approaches, which were categorized according to their being either split or unsplit formulations were made recently in~\cite{hajabdollahi2018symmetrized}, which also developed an efficient and second order accurate method to include sources in the cascaded LBM based on symmetric operator or Strang splitting. In general, the cascaded LB formulation is able to handle low viscous regions or flows at high Reynolds number quite well (see e.g.,~\cite{Geier2015,Ning2015,dubois2015stability,chavez2018improving}) by performing the relaxation process under collision in the moving frame of reference. While naturally preserving the Galilean invariance of the moments independently supported by the lattice, it also leads to the use of higher order terms in the fluid velocity in the equilibria when compared to the standard SRT or MRT formulations thereby contributing to its superior stability characteristics. Furthermore, by tuning the various relaxation times independently, the damping of the non-hydrodynamic kinetic quantities can be achieved~\cite{Ning2015,dubois2015stability,chavez2018improving}.

Complex non-Newtonian fluid flows with nonlinear constitutive relations represent an important class of flows with numerous applications in various scenarios including in engineering, materials and food processing, and geophysical processes~\cite{tanner2000engineering,deville2012mathematical}. The extension of LBM to non-Newtonian fluid flows has received significant  but limited  attention so far, which has been reviewed in~\cite{article}, and some more recent examples of such studies include the works of ~\cite{gabbanelli2005lattice,chai2011multiple,conrad2015accuracy,adam2019numerical}. While most of the prior studies have demonstrated the applications of different LB schemes to 2D non-Newtonian fluid flows, including a recent investigation based on a 2D cascaded LB approach~\cite{adam2019numerical}, very little focus has been given to the development and validation of LBM, particularly with using advanced collision models, for 3D non-Newtonian fluid flows involving well defined benchmark problems. Indeed, due to complexity of handling nonlinear rheological behavior via more general constitutive relations with attendant numerical stability issues and the need for high grid resolutions with relatively higher computational demands, there are only very few studies in 3D even in the context of computational fluid dynamics (CFD) using conventional numerical schemes. For the well-defined problem of 3D non-Newtonian cubic cavity flow, Refs.~\cite{reddy1992finite,elias2006parallel} presented some results at coarse grid resolutions and, recently, Ref.~\cite{jin2017gpu} reported benchmark quality results using a fractional step based finite volume method.

In the present work, we will present a new cascaded LBM in 3D for the simulation of non-Newtonian flows represented by power law fluid based rheology. In this regard, we construct a collision model based on central moments and multiple relaxation times, on a three-dimensional nineteen velocity (D3Q19) lattice. Prior versions of the `cascaded' formulations of the central moment LBM, where the changes in the higher moments depend successively on those preceding lower moments, were developed for Newtonian flows for the D3Q27 or D3Q15 lattice~\cite{Geier2006,premnath2011three}. In this work, we will present a derivation of the cascaded LBM for the D3Q19 lattice, which is a compromise between stability and efficiency, to handle nonlinear constitutive relations. Here, the relaxation times for the second order moments that emulate the nonlinear viscous behavior of the fluid are adjusted by the local shear rate and parametrized by the consistency coefficient and the power law index of the generalized Newtonian (i.e., power law) fluid. Expressions for computing the shear rates based on estimating the components of the strain rate tensor locally via the non-equilibrium moments will be provided for this formulation. The implementation strategy for the sources corresponding to the applied body forces for our 3D cascaded LBM will be according to the operator splitting strategy presented in~\cite{hajabdollahi2018symmetrized}, and will not be discussed further here as the present focus is on developing a modified collision term based on central moments for non-Newtonian flows. However, we will present details on how to implement our cascaded collision formulation, especially some optimization strategies, for efficient LB simulations of 3D non-Newtonian flows. Compared to the conventional schemes for CFD, our 3D LB algorithm is local and is thus naturally suitable for efficient implementation on large scale parallel computers. Moreover, our approach is based on advanced formulation of the collision terms, which is expected to be more stable when compared to other LB models for 3D non-Newtonian flows. We will present a validation study of our 3D cascaded LB scheme for the non-Newtonian flow in a channel, duct and a cubic cavity. In particular, for the latter case involving complex nonlinear fluid motion with power-law rheology in a cubic cavity, we will make direct comparisons of the velocity profiles against the recent benchmark solution~\cite{jin2017gpu} for various Reynolds numbers and power law index magnitudes encompassing both shear thinning and shear thickening fluid behavior. We will also assess the order of accuracy of grid convergence of our new 3D scheme for non-Newtonian flow simulations. Finally, we will present a comparative study of the present 3D cascaded LBM using central moments against the LB schemes based on a SRT model and a MRT formulation based on raw moments for the simulation of non-Newtonian fluid flows with different power law index values and demonstrate improvements in numerical stability achieved by the former when compared to the latter approaches.

It may be noted that recently, Ref.~\cite{chen2020simplified} introduced a so-called simplified LBM for the simulation of non-Newtonian flows and used a 3D cavity flow as a test case. Here, we compare and contrast our approach with this recent work. The method proposed in Ref.~\cite{chen2020simplified} invokes several approximations in formulating the numerical scheme and involves a time-split predictor and corrector steps, which may be subject to splitting errors. Also, the boundary conditions need to be imposed separately for the predictor and corrector steps, with the latter step requiring the use of a linear extrapolation scheme. These may compromise the overall accuracy and mass conservation properties, while our approach is not subjected to these issues as it is based on standard LB discretizations. While we demonstrate our method to be second order accurate in grid convergence (see in  Sec.~\ref{section_numerical_tests}), Ref.~\cite{chen2020simplified} does not present such a numerical convergence study. Moreover, the scheme presented in Ref.~\cite{chen2020simplified} does not have enough additional degrees of freedom to account for additional features, such as the ability to independently adjust the bulk viscosity and shear viscosity to handle more general class of flows, which are naturally accounted for in our approach based on using multiple relaxation times and central moments. In addition, the method relies on using a non-local finite difference scheme to determine the shear rate in implementing the constitutive relation for non-Newtonian fluids, which compromises its parallelization properties; by contrast, our formulation computes the strain rate tensor locally based on non-equilibrium moments for 3D non-Newtonian flows (see Sec.~\ref{sec:Threedimensionalcascaded}). Furthermore, Ref.~\cite{chen2020simplified}, in presenting some results related to 3D cavity flows of non-Newtonian fluids, does not make any comparison with benchmark quality data obtained using well-established Navier-Stokes based solvers, while this work makes such a comparison using the results presented in a recent work~\cite{jin2017gpu}. Finally, unlike this work (see Sec.~\ref{sec:Comparison}), Ref.~\cite{chen2020simplified} does not provide any explicit comparisons of numerical stability with other LB schemes.

This paper is organized as follows. In the next section (Sec.~\ref{governing}), we present an overview of the macroscopic governing equations for generalized non-Newtonian fluid flows and list the attendant non-linear constitutive relation. A detailed exposition of the construction of the 3D cascaded LBM for the D3Q19 lattice is given in Sec.~\ref{sec:Threedimensionalcascaded}. This includes the changes of different moments under cascaded collision and the specification of the local relaxation times for the second order moments parameterized for the power law fluid, where the shear rate is related to the non-equilibrium moments, as well as their implementation strategy. Validation studies involving various canonical non-Newtonian fluid flow problems, including those in a cubic cavity are presented and discussed in Sec.~\ref{section_numerical_tests}. Section~\ref{sec:Comparison} makes numerical stability comparisons of our 3D central moment LBM against a SRT-LBM and a MRT-LBM based on raw moments for the simulation of power law fluid flows in a shear driven problem. In addition, it will present performance data related to computational cost comparisons. The conclusions are finally summarized in Sec.~\ref{sec:summary}.

\section{\label{governing}Governing equations for three-dimensional generalized non-Newtonian fluid flows}
In the present work, we consider the simulation of three-dimensional (3D) generalized Newtonian flows (GNF). The  mathematical model, which is given in terms of the continuity and momentum equations, can be written as
\begin{subequations}
\begin{align}
\frac{\partial{\rho}}{\partial{t}}+\frac{\partial}{\partial{x_j}}{(\rho u_j)}&=0,\label{eq1a}\\
\frac{\partial}{\partial{t}}{(\rho u_i)}+\frac{\partial}{\partial{x_j}}{(\rho u_i u_j)}&=-\frac{\partial{P}}{\partial{x_i}}+\frac{\partial{\sigma_{ij}}}{\partial{x_j}}+F_i, \label{eq1b}
\end{align}
\end{subequations}
where $\rho$ and $u_i$ are the local fluid density and the velocity field, respectively, with $i\in{(x,y,z)}$. Here,
$P$, $F_i$ and $\sigma_{ij}$ represent the pressure, Cartesian component of the imposed body force, and deviatoric viscous stress tensor induced within the GNF, respectively. The viscous stress tensor can be represented as
\begin{equation}
\sigma_{ij} =\mu{(|\dot{\gamma}_{ij}|)}\dot{\gamma}_{ij},
\label{eq2}
\end{equation}
where $\dot{\gamma}_{ij}$ is the shear rate tensor. It is related to the strain rate tensor $S_{ij}$ by
\begin{equation}
\dot{\gamma}_{ij}=2S_{ij}, \qquad  S_{ij}=\frac{1}{2}\left(\partial_{j}u_i+\partial_{i}u_j\right),
\label{eq3}
\end{equation}
and $\mu{(|\dot{\gamma}_{ij}|)}$ is assumed to be an effective or apparent dynamic viscosity of the GNF, where the magnitude of the shear rate $|\dot{\gamma}_{ij}|$ is related to the second invariant of the symmetric strain rate tensor $S_{ij}$ as
\begin{equation}
|\dot{\gamma}_{ij}| =\sqrt{2S_{ij}S_{ij}}.
\label{eq4}
\end{equation}
Here, the usual convention of the summation of the repeated indices is assumed. Thus,
\begin{equation}
|\dot{\gamma}_{ij}|=\sqrt{2\left[S^2_{xx}+S^2_{yy}+S^2_{zz}+2(S^2_{xy}+S^2_{yz}+S^2_{xz})\right]}.\label{eq:shearrateexpression}
\end{equation}
In this study, we consider the constitutive relations for the power law fluid as a representative of a class of GNF for their numerical solution using the 3D cascaded LB method presented in the next section. The constitutive relation for the power law fluid can be described mathematically as follows:
\begin{equation}
\sigma_{ij} =\mu_p{|\dot{\gamma}_{ij}|}^{n-1}\dot{\gamma}_{ij},
\label{eq5}
\end{equation}
where the model parameter $\mu_p$ and the exponent $n$ are known as the consistency coefficient and power law index of the fluid, respectively. Based on the values of the power-law index $n$, the following three different types of non-Newtonian fluids can be represented: $n<1$ corresponds to shear thinning or pseudo-plastic fluid, whereas $n>1$ models shear thickening fluids, and $n=1$ reduces to the Newtonian constitutive relation. Comparing Eqs.~(\ref{eq2}) and (\ref{eq5}), the effective viscosity for the power-law model is given by the following expression
\begin{equation}
\mu_{power}({|\dot{\gamma}_{ij}|}) =\mu_p{|\dot{\gamma}_{ij}|}^{n-1}.
\label{eq6}
\end{equation}

\section{\label{sec:Threedimensionalcascaded}Three-dimensional cascaded LBM for non-Newtonian fluid flows}

\begin{figure}[h]
\begin{center}
\includegraphics[width=.65\textwidth]{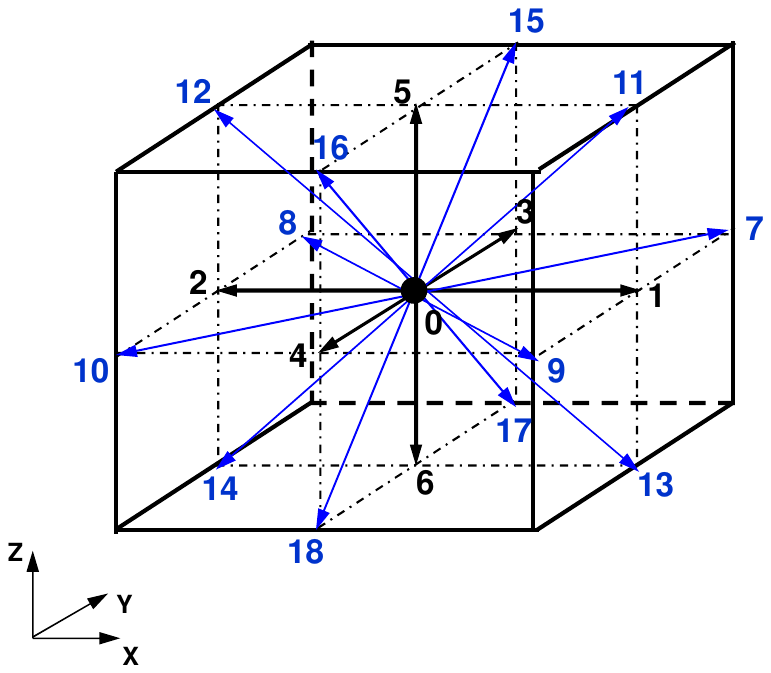}
\caption{\label{fig:d3q19} Three-dimensional, nineteen  particle velocity (D3Q19) lattice.}
\end{center}
\end{figure}

We will formulate our non-Newtonian fluid flow solver based on a cascaded LB approach using the three-dimensional nineteen velocity (D3Q19) lattice, which is shown in Fig.~\ref{fig:d3q19}. The corresponding particle velocity $\bm{e}_{\alpha}$ is represented as
\begin{equation}
\bm{e}_{\alpha} = \left\{\begin{array}{ll}
   {(0,0,0),}&{\alpha=0}\\
   {(\pm 1,0,0), (0,\pm 1,0), (0,0,\pm 1),}&{\alpha=1,\cdots,6}\\
   {(\pm 1,\pm 1,0), (\pm 1,0,\pm 1), (0,\pm 1,\pm 1),}&{\alpha=7,\cdots,18}
\end{array} \right.
\label{eq:velocityd3qq19}
\end{equation}
In this work, the Greek and Latin subscripts are used for the particle velocity directions and Cartesian coordinate directions, respectively. The bare raw moments in the LBM are defined in terms of the distribution function $f_{\alpha}$ as $\sum_{\alpha=0}^{18}e_{\alpha x}^m e_{\alpha y}^n e_{\alpha z}^pf_{\alpha}$, where $m$, $n$ and $p$ are integers, and $(m+n+p)$ represents the order of the corresponding moment component. We use the Dirac's bra-ket notation in this paper to denote the basis vectors. For example, if $\bra{\mathbf{a}}$ and $\ket{\mathbf{b}}$ represents a row vector and a column vector, respectively, then their inner product is given by $\braket{\mathbf{a}|\mathbf{b}}$. Now, the nineteen non-orthogonal basis vectors based on the combination of the  monomials $e_{\alpha x}^m e_{\alpha y}^n e_{\alpha z}^p$ in an ascending order supported by the D3Q19 lattice can be listed as follows:
\allowdisplaybreaks
\begin{eqnarray}
\ket{T_{0}}&=&\ket{\mathbf{1}},\quad \ket{T_{1}}=\ket{\mathbf{e_{x}}},\quad \ket{T_{2}}=\ket{\mathbf{e_{y}}},\quad \ket{T_{3}}=\ket{\mathbf{e_{z}}},\nonumber\\
\ket{T_{4}}&=&\ket{\mathbf{e_{x}}\mathbf{e_{y}}},\quad \ket{T_{5}}=\ket{\mathbf{e_{x}}\mathbf{e_{z}}},\quad \ket{T_{6}}=\ket{\mathbf{e_{y}}\mathbf{e_{z}}},\nonumber\\
\ket{T_{7}}&=&\ket{\mathbf{e_{x}}^2-\mathbf{e_{y}}^2},\quad \ket{T_{8}}=\ket{\mathbf{e_{x}}^2-\mathbf{e_{z}}^2},\quad \ket{T_{9}}=\ket{\mathbf{e_{x}}^2+\mathbf{e_{y}}^2+\mathbf{e_{z}}^2},\nonumber\\
\ket{T_{10}}&=&\ket{\mathbf{e_{x}}\left(\mathbf{e_{x}}^2+\mathbf{e_{y}}^2+\mathbf{e_{z}}^2\right)},\quad
\ket{T_{11}}=\ket{\mathbf{e_{y}}\left(\mathbf{e_{x}}^2+\mathbf{e_{y}}^2+\mathbf{e_{z}}^2\right)},\nonumber\\
\ket{T_{12}}&=&\ket{\mathbf{e_{z}}\left(\mathbf{e_{x}}^2+\mathbf{e_{y}}^2+\mathbf{e_{z}}^2\right)},\nonumber\\
\ket{T_{13}}&=&\ket{\mathbf{e_{x}}\left(\mathbf{e_{y}}^2-\mathbf{e_{z}}^2\right)},\quad
\ket{T_{14}}=\ket{\mathbf{e_{y}}\left(\mathbf{e_{z}}^2-\mathbf{e_{x}}^2\right)},\quad
\ket{T_{15}}=\ket{\mathbf{e_{z}}\left(\mathbf{e_{x}}^2-\mathbf{e_{y}}^2\right)},\nonumber\\
\ket{T_{16}}&=&\ket{\mathbf{e_{x}}^2\mathbf{e_{y}}^2+\mathbf{e_{x}}^2\mathbf{e_{z}}^2+\mathbf{e_{y}}^2\mathbf{e_{z}}^2},\quad
\ket{T_{17}}=\ket{\mathbf{e_{x}}^2\mathbf{e_{y}}^2+\mathbf{e_{x}}^2\mathbf{e_{z}}^2-\mathbf{e_{y}}^2\mathbf{e_{z}}^2},\nonumber\\
\ket{T_{18}}&=&\ket{\mathbf{e_{x}}^2\mathbf{e_{y}}^2-\mathbf{e_{x}}^2\mathbf{e_{z}}^2}.\nonumber
\end{eqnarray}
Here, the components of the basis vectors $\ket{\mathbf{1}},\ket{\mathbf{e_{x}}},\ket{\mathbf{e_{y}}}$ and $\ket{\mathbf{e_{z}}}$ corresponding to the conserved moments based on which the moment basis can be constructed are given by
\begin{eqnarray}
\ket{\mathbf{1}} &=&\left(1,1,1,1,1,1,1,1,1,1,1,1,1,1,1,1,1,1,1\right)^\dagger, \nonumber \\
\ket{\mathbf{e_{x}}} &=&\left(0,1,-1,0,0,0,0,1,-1,1,-1,1,-1,1,-1,0,0,0,0\right)^\dagger, \nonumber \\
\ket{\mathbf{e_{y}}} &=&\left(0,0,0,1,-1,0,0,1,1,-1,-1,0,0,0,0,1,-1,1,-1\right)^\dagger,\nonumber \\
\ket{\mathbf{e_{z}}} &=&\left(0,0,0,0,0,1,-1,0,0,0,0,1,1,-1,-1,1,1,-1,-1\right)^\dagger,\nonumber
\end{eqnarray}
where $\dagger$ is the transpose operator. As in our previous work~\cite{premnath2011three}, in order to formulate a `cascaded' LB formulation, where the changes in higher order moments successively depend on those of the preceding lower order ones, for the D3Q19 lattice, the above $\ket{T_{j}}$ vectors, where $j=0,1,2,..,18,$ can be transformed into the following equivalent set of orthogonal basis vectors via the standard Gram-Schmidt procedure, which then read as
\begin{eqnarray}
\ket{K_{0}}&=&\ket{\mathbf{1}},\quad
\ket{K_{1}}=\ket{\mathbf{e_{x}}},\quad
\ket{K_{2}}=\ket{\mathbf{e_{y}}},\quad
\ket{K_{3}}=\ket{\mathbf{e_{z}}},\nonumber\\
\ket{K_{4}}&=&\ket{\mathbf{e_{x}}\mathbf{e_{y}}},\quad
\ket{K_{5}}=\ket{\mathbf{e_{x}}\mathbf{e_{z}}},\quad
\ket{K_{6}}=\ket{\mathbf{e_{y}}\mathbf{e_{z}}},\nonumber\\
\ket{K_{7}}&=&\ket{\mathbf{e_{x}}^2-\mathbf{e_{y}}^2},\quad
\ket{K_{8}}=\ket{\mathbf{e_{x}}^2+\mathbf{e_{y}}^2+\mathbf{e_{z}}^2}-3\ket{\mathbf{e_{z}}^2},\nonumber\\
\ket{K_{9}}&=&19\ket{\mathbf{e_{x}}^2+\mathbf{e_{y}}^2+\mathbf{e_{z}}^2}-30\ket{\mathbf{1}},\nonumber\\
\ket{K_{10}}&=&5\ket{\mathbf{e_{x}}\left(\mathbf{e_{x}}^2+\mathbf{e_{y}}^2+\mathbf{e_{z}}^2\right)}-9\ket{\mathbf{e_{x}}},\quad
\ket{K_{11}}=5\ket{\mathbf{e_{y}}\left(\mathbf{e_{x}}^2+\mathbf{e_{y}}^2+\mathbf{e_{z}}^2\right)}-9\ket{\mathbf{e_{y}}},\nonumber\\
\ket{K_{12}}&=&5\ket{\mathbf{e_{z}}\left(\mathbf{e_{x}}^2+\mathbf{e_{y}}^2+\mathbf{e_{z}}^2\right)}-9\ket{\mathbf{e_{z}}},\nonumber\\
\ket{K_{13}}&=&\ket{\mathbf{e_{x}}\mathbf{e_{y}}^2-\mathbf{e_{x}}\mathbf{e_{z}}^2},\quad
\ket{K_{14}}=\ket{\mathbf{e_{z}}^2\mathbf{e_{y}}-\mathbf{e_{y}}\mathbf{e_{x}^2}},\quad
\ket{K_{15}}=\ket{\mathbf{e_{x}}^2\mathbf{e_{z}}-\mathbf{e_{y}}^2\mathbf{e_{z}}},\nonumber\\
\ket{K_{16}}&=&21\ket{\mathbf{e_{x}}^2\mathbf{e_{y}}^2+\mathbf{e_{x}}^2\mathbf{e_{z}}^2+\mathbf{e_{y}}^2\mathbf{e_{z}}^2}-16\ket{\mathbf{e_{x}}^2+\mathbf{e_{y}}^2+\mathbf{e_{z}}^2}+12\ket{\mathbf{1}},\nonumber\\
\ket{K_{17}}&=&3\ket{\mathbf{e_{x}}^2\mathbf{e_{y}}^2+\mathbf{e_{x}}^2\mathbf{e_{z}}^2-2\mathbf{e_{y}}^2\mathbf{e_{z}}^2}-2\ket{2\mathbf{e_{x}}^2-\mathbf{e_{y}}^2-\mathbf{e_{z}}^2},\nonumber\\
\ket{K_{18}}&=&3\ket{\mathbf{e_{x}}^2\mathbf{e_{y}}^2-\mathbf{e_{x}}^2\mathbf{e_{z}}^2}-2\ket{\mathbf{e_{y}}^2-\mathbf{e_{z}}^2}.\nonumber
\end{eqnarray}
These vectors can be used to form an orthogonal matrix $\tensor{K}$ that maps changes in moments under cascaded collision as
\begin{equation}
\tensor{K}=\left[\ket{K_{0}},\ket{K_{1}},\ket{K_{2}},\ldots,\ket{K_{18}}\right]
\label{eq:collisionmatrix1}
\end{equation}
whose elements are given by
\newline
$\tensor{K}=$
\scriptsize
$
\left(
\begin{array}{rrrrrrrrrrrrrrrrrrrrrrrrrrr}
1	&0	&0	&0	&0	&0	&0	&0	&0	&-30	&0	&0	&0	&0	&0	&0	&12	&0	&0	\\
1	&1	&0	&0	&0	&0	&0	&1	&1 &-11 &-4	&0	&0	&0	&0	&0	&-4	&-4	&0\\
1	&-1	&0	&0	&0	&0	&0	&1	&1	&-11	&4	&0	&0	&0	&0	&0	&-4	&-4	&0	\\
1	&0	&1	&0	&0	&0	&0	&-1	&1	&-11	&0	&-4	&0	&0	&0	&0	&-4		&2	&-2	\\
1	&0	&-1	&0	&0	&0	&0	&-1	&1	&-11	&0	&4	&0	&0	&0	&0	&-4	&2	&-2\\
1	&0	&0	&1	&0	&0	&0	&0	&-2	&-11	&0	&0	&-4	&0	&0	&0	&-4	&2	&2\\
1	&0	&0	&-1	&0	&0	&0	&0	&-2	&-11	&0	&0	&4	&0	&0	&0	&-4	&2	&2\\
1	&1	&1	&0	&1	&0	&0	&0	&2	&8	&1	&1	&0	&1	&-1	&0	&1	&1	&1\\
1	&-1	&1	&0	&-1	&0	&0	&0	&2	&8	&-1	&1	&0	&-1	&-1	&0	&1	&1	&1\\
1	&1	&-1	&0	&-1	&0	&0	&0	&2	&8	&1	&-1	&0	&1	&1	&0	&1	&1	&1\\
1	&-1	&-1	&0	&1	&0	&0	&0	&2	&8	&-1	&-1	&0	&-1	&1	&0	&1	&1	&1\\
1	&1	&0	&1	&0	&1	&0	&1	&-1	&8	&1	&0	&1	&-1	&0	&1	&1	&1	&-1\\
1	&-1	&0	&1	&0	&-1	&0	&1	&-1	&8	&-1	&0	&1	&1	&0	&1	&1	&1	&-1\\
1	&1	&0	&-1	&0	&-1	&0	&1	&-1	&8	&1	&0	&-1	&-1	&0	&-1	&1	&1	&-1\\
1	&-1	&0	&-1	&0	&1	&0	&1	&-1	&8	&-1	&0	&-1	&1	&0	&-1	&1	&1	&-1\\
1	&0	&1	&1	&0	&0	&1	&-1	&-1	&8	&0	&1	&1	&0	&-1	&-1	&1	&-2	&0\\
1	&0	&-1	&1	&0	&0	&-1	&-1	&-1	&8	&0	&-1	&1	&0	&-1	&-1	&1	&-2	&0\\
1	&0	&1	&-1	&0	&0	&-1	&-1	&-1	&8	&0	&1	&-1	&0	&1	&1	&1	&-2	&0\\
1	&0	&-1	&-1	&0	&0	&1	&-1	&-1	&8	&0	&-1	&-1	&0	&-1	&1	&1	&-2	&0\\
\end{array}
\right)
$
\normalsize \newline

Next, we define the continuous central moment of the particle distribution function $f$ and its attractor or the local equilibriuam $f^{at}$  shifted by the local fluid velocity of order $(m+n+p)$ as
\begin{eqnarray}
\left[\begin{array}{l}
{{\hat \Pi}_{{x^m}{y^n}{z^p}}}\\
{{\hat \Pi}^{at}_{{x^m}{y^n}{z^p}}}
\end{array} \right]=\int_{-\infty}^{\infty}\int_{-\infty}^{\infty}\int_{-\infty}^{\infty}
{\left[ \begin{array}{l}
{f }\\
f^{at}\\
\end{array} \right]}
(\xi_x-u_x)^m(\xi_y-u_y)^n(\xi_z-u_z)^pd\xi_xd\xi_yd\xi_z.
\label{eq:centralmomentfdefinition}
\end{eqnarray}
Here, and in the rest of this paper, we employ "hat" over a symbol to represent the values in the space of moments. One possibility is to consider the local Maxwell-Boltzmann distribution function in the continuous particle velocity space ($\bm{\xi}=(\xi_x$,$\xi_y$,$\xi_z)$) as the attractor
\begin{equation}
f^\mathcal{M}\equiv
f^\mathcal{M}(\rho,\bm{u}; \bm{\xi})=\frac{\rho}{2\pi c_s^{3/2}}\exp\left[-\left(\bm{\xi}-\bm{u}\right)^2/(2c_s^2)\right],\label{eq:Maxwellian}
\end{equation}
where we typically choose $c_s^2=1/3$, and its central moments follow via the definition given in Eq.~(\ref{eq:centralmomentfdefinition}), which can be
written as follows:
\begin{eqnarray}
\widehat{\Pi}^{\mathcal{M}}_{0}&=&\rho, \nonumber\\
\widehat{\Pi}^{\mathcal{M}}_{i}&=&0, \nonumber\\
\widehat{\Pi}^{\mathcal{M}}_{ii}&=&c_s^2\rho,\nonumber\\
\widehat{\Pi}^{\mathcal{M}}_{ij}&=&0,\quad i \neq j,\nonumber\\
\widehat{\Pi}^{\mathcal{M}}_{ijj}&=&0,\quad i \neq j,\nonumber\\
\widehat{\Pi}^{\mathcal{M}}_{ijk}&=&0,\quad i \neq j \neq k,\nonumber\\
\widehat{\Pi}^{\mathcal{M}}_{iijj}&=&c_s^4\rho,\quad i \neq j.
\end{eqnarray}
In the above equation and the one that follows, for ease of representation, we do not assume the summation over the repeated indices. More generally, as suggested in Ref.~\cite{geier2009factorized}, we consider the factorized form of attractors that guide the collision process, i.e.,
\begin{eqnarray}
\widehat{\Pi}^{at}_{i}&=&\widetilde{\widehat{\Pi}}_{i}=0, \nonumber\\
\widehat{\Pi}^{at}_{ij}&=&\widetilde{\widehat{\Pi}}_{i}\widetilde{\widehat{\Pi}}_{j}=0, \quad i \neq j \nonumber\\
\widehat{\Pi}^{at}_{iij}&=&\widetilde{\widehat{\Pi}}_{ii}\widetilde{\widehat{\Pi}}_{j}=0, \nonumber\\
\widehat{\Pi}^{at}_{ijk}&=&\widetilde{\widehat{\Pi}}_{i}\widetilde{\widehat{\Pi}}_{j}\widetilde{\widehat{\Pi}}_{k}=0,\nonumber\\
\widehat{\Pi}^{at}_{iijj}&=&\widetilde{\widehat{\Pi}}_{ii}\widetilde{\widehat{\Pi}}_{jj},
\end{eqnarray}
where `tilde' denote the post-collision values and $\widetilde{\widehat{\Pi}}_{x^my^nz^p}$ represents the central moments of order $(m+n+p)$ of the post-collision distribution function $\widetilde{f}$:
\begin{equation}
\widetilde{\widehat{\Pi}}_{x^my^nz^p}=\int_{-\infty}^{\infty}\int_{-\infty}^{\infty}\int_{-\infty}^{\infty}\widetilde{f}(\xi_x-u_x)^m(\xi_y-u_y)^n(\xi_z-u_z)^pd\xi_xd\xi_yd\xi_z.
\label{eq:centralmomentfdefinitionpostcollision}
\end{equation}
Nevertheless, the diagonal components of the lower second-order central moments of the attractor should be obtained from the Maxwellian central moments, i.e., $\widehat{\Pi}^{\mathcal{M}}_{ii}=c_s^2\rho$  in order to correctly recover the pressure and momentum flux in the macroscopic flow equations~\cite{geier2009factorized}. Then, the $19$ independent components of the central moment attractors based on the factorized formulation
can be written as
\begin{eqnarray}
\widehat{\Pi}^{at}_{0}&=&0,\widehat{\Pi}^{at}_{x}=\widehat{\Pi}^{at}_{y}=\widehat{\Pi}^{at}_{z}=0, \nonumber\\
\widehat{\Pi}^{at}_{xx}&=&\widehat{\Pi}^{at}_{yy}=\widehat{\Pi}^{at}_{zz}=c_s^2\rho, \nonumber\\
\widehat{\Pi}^{at}_{xy}&=&\widehat{\Pi}^{at}_{xz}=\widehat{\Pi}^{at}_{yz}=0,\nonumber\\
\widehat{\Pi}^{at}_{xyy}&=&\widehat{\Pi}^{at}_{xzz}=\widehat{\Pi}^{at}_{xxy}=\widehat{\Pi}^{at}_{yzz}=\widehat{\Pi}^{at}_{xxz}=\widehat{\Pi}^{at}_{yyz}=\widehat{\Pi}^{at}_{xyz}=0,\nonumber\\
\widehat{\Pi}^{at}_{xxyy}&=&\widetilde{\widehat{\Pi}}_{xx}\widetilde{\widehat{\Pi}}_{yy},\nonumber\\
\widehat{\Pi}^{at}_{xxzz}&=&\widetilde{\widehat{\Pi}}_{xx}\widetilde{\widehat{\Pi}}_{zz},\nonumber\\
\widehat{\Pi}^{at}_{yyzz}&=&\widetilde{\widehat{\Pi}}_{yy}\widetilde{\widehat{\Pi}}_{zz}.
\end{eqnarray}

Based on the above, we will now construct a 3D cascaded LBM by defining a discrete distribution function supported by the discrete particle velocity set $\bm{e}_{\alpha}$ as $\mathbf{f}=(f_0,f_1,f_2,\ldots,f_{18})^\dagger$ and its collision term as
$\mathbf{\Omega}^{c}=\ket{\Omega_{\alpha}^{c}}=(\Omega_{0}^{c},\Omega_{1}^{c},\Omega_{2}^{c},\ldots,\Omega_{18}^{c})^\dagger$, which will be based on the relaxation of various central moments to their attractors. The cascaded LBE results from a special discretization of the continuous Boltzmann equation by integrating along the particle characteristics as
\begin{equation}
f_{\alpha}(\bm{x}+\bm{e}_{\alpha},t+1)=f_{\alpha}(\bm{x},t)+\Omega^{c}_{\alpha}(\bm{x},t)
\label{eq:cascadedLBE1}
\end{equation}
Here, the collision term $\Omega_{\alpha}^{c}$ is a function of the unknown changes in various moments under collision  $\mathbf{\widehat{g}}=(\widehat{g}_0,\widehat{g}_1,\widehat{g}_2,\ldots,\widehat{g}_{18})^\dagger$ resulting from the state $\mathbf{f}$, which are then mapped to the changes in the distribution functions via the orthogonal mapping matrix $\tensor{K}$ as~\cite{Geier2006}
\begin{equation}
\Omega_{\alpha}^{c}\equiv \Omega_{\alpha}^{c}(\mathbf{f},\mathbf{\widehat{g}})=(\tensor{K}\cdot\mathbf{\widehat{g}})_{\alpha},
\label{eq:cascadecollision1}
\end{equation}
For convenience, the above cascaded LBE (Eq.~(\ref{eq:cascadedLBE1})) can reformulated as an algorithm in the form of the usual collision and streaming steps, respectively, as
\begin{eqnarray}
\widetilde{f}_{\alpha}(\bm{x},t)&=&f_{\alpha}(\bm{x},t)+\Omega^{c}_{\alpha}{(\bm{x},t)},\label{eq:cascadedcollision1}\\
f_{\alpha}(\bm{x}+\bm{e}_{\alpha},t+1)&=&\widetilde{f}_{\alpha}(\bm{x},t).\label{eq:cascadedstreaming1}
\end{eqnarray}
The fluid dynamical variables, i.e., the local density and momentum, and the pressure field $P$ are determined from the updated distribution functions as
\begin{subequations}
\begin{eqnarray}
\rho&=&\sum_{\alpha=0}^{18}f_{\alpha}=\braket{\mathbf{f}|\mathbf{1}},\quad P = \rho c_s^2\label{eq:hydrodynamicfields1}\\
\rho u_i&=&\sum_{\alpha=0}^{18}f_{\alpha} e_{\alpha i}=\braket{\mathbf{f}|\mathbf{e}_i}, i \in {x,y,z}.\label{eq:hydrodynamicfields2}
\end{eqnarray}
\end{subequations}

In order to derive the expressions for the change of moments under collision $\widehat{\mathbf{g}}$ for the cascaded LBM, we need to also define the following \emph{discrete} central moments of the distribution functions $\mathbf{f}=(f_0,f_1,f_2,\ldots,f_{18})^\dagger$, their attractors $\mathbf{f^{at}}=(f_0^{at},f_1^{at},f_2^{at},\ldots,f_{18}^{at})^\dagger$, and the post-collision distribution functions~$\mathbf{\widetilde{f}}=(\widetilde{f}_0,\widetilde{f}_1,\widetilde{f}_2,\ldots,\widetilde{f}_{18})^\dagger$:
\begin{eqnarray}
\widehat{\kappa}_{x^m y^n z^p}&=&\braket{(\mathbf{e_{x}}-u_x\mathbf{1})^m(\mathbf{e_{y}}-u_y\mathbf{1})^n(\mathbf{e_{z}}-u_z\mathbf{1})^p|\mathbf{f}},\label{eq:centralmomentdistributionfunction1}\nonumber\\
\widehat{\kappa}_{x^m y^n z^p}^{at}&=&\braket{(\mathbf{e_{x}}-u_x\mathbf{1})^m(\mathbf{e_{y}}-u_y\mathbf{1})^n(\mathbf{e_{z}}-u_z\mathbf{1})^p|\mathbf{f^{at}}},\label{eq:centralmomentMaxwelldistribution1}\nonumber\\
\widetilde{\widehat{\kappa}}_{x^m y^n z^p}&=&\braket{(\mathbf{e_{x}}-u_x\mathbf{1})^m(\mathbf{e_{y}}-u_y\mathbf{1})^n(\mathbf{e_{z}}-u_z\mathbf{1})^p|\mathbf{\widetilde{f}}},\label{eq:postcentralmomentdistributionfunction1}.
\end{eqnarray}
For achieving highest possible accuracy, we match the discrete central moments of the attractors supported by the D3Q19 lattice with the corresponding continuous central moments, i.e., $\widehat{\kappa}_{x^m y^n z^p}^{at}=\widehat{\Pi}^{at}_{x^m y^n z^p}$. As a result, the discrete central moments of the attractors can be listed as follows:
\begin{eqnarray}                                                       \widehat{\kappa}^{at}_{0}&=&0,\widehat{\kappa}^{at}_{x}=\widehat{\kappa}^{at}_{y}=\widehat{\kappa}^{at}_{z}=0, \nonumber\\
\widehat{\kappa}^{at}_{xx}&=&\widehat{\kappa}^{at}_{yy}=\widehat{\kappa}^{at}_{zz}=c_s^2\rho, \nonumber\\
\widehat{\kappa}^{at}_{xy}&=&\widehat{\kappa}^{at}_{xz}=\widehat{\kappa}^{at}_{yz}=0,\nonumber\\
\widehat{\kappa}^{at}_{xyy}&=&\widehat{\kappa}^{at}_{xzz}=\widehat{\kappa}^{at}_{xxy}=\widehat{\kappa}^{at}_{yzz}=\widehat{\kappa}^{at}_{xxz}=\widehat{\kappa}^{at}_{yyz}=\widehat{\kappa}^{at}_{xyz}=0,\nonumber\\
\widehat{\kappa}^{at}_{xxyy}&=&\widetilde{\widehat{\kappa}}_{xx}\widetilde{\widehat{\kappa}}_{yy},\nonumber\\
\widehat{\kappa}^{at}_{xxzz}&=&\widetilde{\widehat{\kappa}}_{xx}\widetilde{\widehat{\kappa}}_{zz},\nonumber\\
\widehat{\kappa}^{at}_{yyzz}&=&\widetilde{\widehat{\kappa}}_{yy}\widetilde{\widehat{\kappa}}_{zz}.\label{eq:discretecentralmomentattractors}
\end{eqnarray}
We also require the raw moments in order to relate to the results obtained in terms of the central moments in the construction of the cascaded collision term in what follows, which we define as
\begin{equation}
\widehat{\kappa}_{x^m y^n z^p}^{'}=\braket{\mathbf{e_{x}}^m\mathbf{e_{y}}^n\mathbf{e_{z}}^p|\mathbf{f}},\label{eq:rawmomentdistributionfunction1}
\end{equation}
where the superscript `prime' ($'$) is used for the various raw moments in order to distinguish the raw moments from the central moments that are designated without the primes.

\subsection{\label{sec:cascadedcollisionexpressions}Change of moments under cascaded collision for D3Q19 lattice}
We now need the various moments of the cascaded collision term, i.e.,
\begin{equation}
\braket{(\tensor{K}\cdot \mathbf{\widehat{g}})|\mathbf{e_{x}}^m \mathbf{e_{y}}^n \mathbf{e_{z}}^p}
\end{equation}
to proceed further. Since the zeroth and first order moments are collision invariants, it follows that the corresponding moment changes under collision are zero, i.e., $\widehat{g}_{0}=\widehat{g}_{1}=\widehat{g}_{2}=\widehat{g}_{3}=0$. In view of the fact that $\tensor{K}$ is orthogonal, the expressions for the various non-conserved components can be readily evaluated, which read as
\begin{eqnarray*}
\braket{(\tensor{K}\cdot \mathbf{\widehat{g}})|\mathbf{e_{x}}\mathbf{e_{y}}}&=&4\widehat{g}_{4},\nonumber\\
\braket{(\tensor{K}\cdot \mathbf{\widehat{g}})|\mathbf{e_{x}}\mathbf{e_{z}}}&=&4\widehat{g}_{5},\nonumber\\
\braket{(\tensor{K}\cdot \mathbf{\widehat{g}})|\mathbf{e_{y}}\mathbf{e_{z}}}&=&4\widehat{g}_{6},\nonumber\\
\braket{(\tensor{K}\cdot \mathbf{\widehat{g}})|\mathbf{e_{x}}^2}&=&6\widehat{g}_{7}+6\widehat{g}_{8}+42\widehat{g}_{9},\nonumber\\
\braket{(\tensor{K}\cdot \mathbf{\widehat{g}})|\mathbf{e_{y}}^2}&=&-6\widehat{g}_{7}+6\widehat{g}_{8}+42\widehat{g}_{9},\nonumber\\
\braket{(\tensor{K}\cdot \mathbf{\widehat{g}})|\mathbf{e_{z}}^2}&=&-12\widehat{g}_{8}+42\widehat{g}_{9},\nonumber\\
\braket{(\tensor{K}\cdot \mathbf{\widehat{g}})|\mathbf{e_{x}}\mathbf{e_{y}}^2}&=&4\widehat{g}_{10}+4\widehat{g}_{13},\nonumber\\
\braket{(\tensor{K}\cdot \mathbf{\widehat{g}})|\mathbf{e_{x}}\mathbf{e_{z}}^2}&=&4\widehat{g}_{10}-4\widehat{g}_{13},\nonumber\\
\braket{(\tensor{K}\cdot \mathbf{\widehat{g}})|\mathbf{e_{x}}^2\mathbf{e_{y}}}&=&4\widehat{g}_{11}-4\widehat{g}_{14},\nonumber\\
\braket{(\tensor{K}\cdot \mathbf{\widehat{g}})|\mathbf{e_{y}}\mathbf{e_{z}}^2}&=&4\widehat{g}_{11}+4\widehat{g}_{14},\nonumber\\
\braket{(\tensor{K}\cdot \mathbf{\widehat{g}})|\mathbf{e_{x}}^2\mathbf{e_{z}}}&=&4\widehat{g}_{12}+4\widehat{g}_{15},\nonumber\\
\braket{(\tensor{K}\cdot \mathbf{\widehat{g}})|\mathbf{e_{y}}^2\mathbf{e_{z}}}&=&4\widehat{g}_{12}-4\widehat{g}_{15},\nonumber\\
\braket{(\tensor{K}\cdot \mathbf{\widehat{g}})|\mathbf{e_{x}}^2\mathbf{e_{y}}^2}&=&8\widehat{g}_{8}+32\widehat{g}_{9}+4\widehat{g}_{16}+4\widehat{g}_{17}+4\widehat{g}_{18},\nonumber\\
\braket{(\tensor{K}\cdot \mathbf{\widehat{g}})|\mathbf{e_{x}}^2\mathbf{e_{z}}^2}&=&4\widehat{g}_{7}-4\widehat{g}_{8}+32\widehat{g}_{9}+4\widehat{g}_{16}+4\widehat{g}_{17}-4\widehat{g}_{18},\nonumber\\
\braket{(\tensor{K}\cdot \mathbf{\widehat{g}})|\mathbf{e_{y}}^2\mathbf{e_{z}}^2}&=&-4\widehat{g}_{7}-4\widehat{g}_{8}+32\widehat{g}_{9} +4\widehat{g}_{16}-8\widehat{g}_{17}.\label{eq:momentscascadedcollisionterm}
\end{eqnarray*}

Then, the expressions for the change of different moments under collision  $\mathbf{\widehat{g}}$ of the 3D cascaded LBE appearing in the term $\tensor{K}\cdot \mathbf{\widehat{g}}$ can be derived as follows. In essence, the procedure starts from the lowest order, non-conserved, central moments (i.e., $\widehat{\kappa}_{xy}$, $\widehat{\kappa}_{xz}$, $\widehat{\kappa}_{yz}$ and higher), and their post-collision central moments (i.e., $\widetilde{\widehat{\kappa}}_{xy}, \widetilde{\widehat{\kappa}}_{xz}$, $\widetilde{\widehat{\kappa}}_{yz}$ and higher) are successively set equal to the corresponding attractors given in Eq.~(\ref{eq:discretecentralmomentattractors}) (i.e., $\widehat{\kappa}_{xy}^{at}, \widehat{\kappa}_{xz}^{at}$ and $\widehat{\kappa}_{yz}^{at}$, and higher). This intermediate step can provide tentative expressions for $\mathbf{\widehat{g}}$ based on an equilibrium assumption, which are then modified to allow for collision as a relaxation process. They are multiplied with a corresponding relaxation parameter that results in the final expressions for the change of moments under collision  $\widehat{g}_{\alpha}$ for a given order~\cite{Geier2006}. In this step, the relaxation parameter needs to multiply only with those terms that are not yet in post-collision states. In other words, the relaxation of different central moments to their corresponding attractors may be formally represented as
\begin{equation}
\braket{(\mathbf{e_{x}}-u_x\mathbf{1})^m(\mathbf{e_{y}}-u_y\mathbf{1})^n(\mathbf{e_{z}}-u_z\mathbf{1})^p|(\tensor{K}\cdot \mathbf{\widehat{g}})}=\omega_*\left( \widehat{\kappa}_{x^my^nz^p}^{at}-\widehat{\kappa}_{x^my^nz^p} \right),\label{eq:centralmomentrelaxation}
\end{equation}
where $\omega_*$ represents the relaxation parameter for the central moment of order $(m+n+p)$. In simplifying the above equation (Eq.~(\ref{eq:centralmomentrelaxation})) for different combinations of $m$, $n$ and $p$ representing the independent moments supported by the D3Q19 lattice, we use Eq.~(\ref{eq:momentscascadedcollisionterm}), which then lead to a cascaded structure, i.e., the change of moments under cascaded collision depend successively on those of the preceding lower order moment changes. The final results, after transforming the central moments to raw moments (see Eq.~(\ref{eq:rawmomentdistributionfunction1}) for definition) for convenience, can then be summarized as follows:
\begin{equation}
\widehat{g}_4=\frac{\omega_4}{4}\left[-\widehat{\kappa}_{xy}^{'}+\rho u_xu_y\right],\label{eq:collisionkernelg4}
\end{equation}
\begin{equation}
\widehat{g}_5=\frac{\omega_5}{4}\left[-\widehat{\kappa}_{xz}^{'}+\rho u_xu_z\right],\label{eq:collisionkernelg5}
\end{equation}
\begin{equation}
\widehat{g}_6=\frac{\omega_6}{4}\left[-\widehat{\kappa}_{yz}^{'}+\rho u_yu_z\right].\label{eq:collisionkernelg6}
\end{equation}
\begin{equation}
\widehat{g}_7=\frac{\omega_7}{12}\left[-(\widehat{\kappa}_{xx}^{'}-\widehat{\kappa}_{yy}^{'})+\rho (u_x^2-u_y^2)\right],\label{eq:collisionkernelg7}
\end{equation}
\begin{eqnarray}
\widehat{g}_8&=&\frac{\omega_8}{36}\left[-(\widehat{\kappa}_{xx}^{'}+\widehat{\kappa}_{yy}^{'}-2\widehat{\kappa}_{zz}^{'})+
\rho (u_x^2+u_y^2-2u_z^2)\right],\label{eq:collisionkernelg8}
\end{eqnarray}
\begin{eqnarray}
\widehat{g}_9&=&\frac{\omega_9}{126}\left[-(\widehat{\kappa}_{xx}^{'}+\widehat{\kappa}_{yy}^{'}+\widehat{\kappa}_{zz}^{'})+
\rho (u_x^2+u_y^2+u_z^2)+\rho\right].\label{eq:collisionkernelg9}
\end{eqnarray}
\begin{eqnarray}
\widehat{g}_{10}&=&\frac{\omega_{10}}{8}\left[-(\widehat{\kappa}_{xyy}^{'}+\widehat{\kappa}_{xzz}^{'})+
2(u_y\widehat{\kappa}_{xy}^{'}+u_z\widehat{\kappa}_{xz}^{'})+
u_x(\widehat{\kappa}_{yy}^{'}+\widehat{\kappa}_{zz}^{'})\right.\nonumber\\
&&\left.-2\rho u_x(u_y^2+u_z^2)\right]+(u_y\widehat{g}_4+u_z\widehat{g}_5)+\frac{3}{4}u_x(-(\widehat{g}_7+\widehat{g}_8)+14\widehat{g}_9),\label{eq:collisionkernelg10}
\end{eqnarray}
\begin{eqnarray}
\widehat{g}_{11}&=&\frac{\omega_{11}}{8}\left[-(\widehat{\kappa}_{yzz}^{'}+\widehat{\kappa}_{xxy}^{'})+
2(u_x\widehat{\kappa}_{xy}^{'}+u_z\widehat{\kappa}_{yz}^{'})+
u_y(\widehat{\kappa}_{xx}^{'}+\widehat{\kappa}_{zz}^{'})\right.\nonumber\\
&&\left.-2\rho u_y(u_x^2+u_z^2)\right]+(u_x\widehat{g}_4+u_z\widehat{g}_6)+\frac{3}{4}u_y(\widehat{g}_7-\widehat{g}_8+14\widehat{g}_9),\label{eq:collisionkernelg11}
\end{eqnarray}
\begin{eqnarray}
\widehat{g}_{12}&=&\frac{\omega_{12}}{8}\left[-(\widehat{\kappa}_{xxz}^{'}+\widehat{\kappa}_{yyz}^{'})+
2(u_x\widehat{\kappa}_{xz}^{'}+u_y\widehat{\kappa}_{yz}^{'})+
u_z(\widehat{\kappa}_{xx}^{'}+\widehat{\kappa}_{yy}^{'})\right.\nonumber\\
&&\left.-2\rho u_z(u_x^2+u_y^2)\right]+(u_x\widehat{g}_5+u_y\widehat{g}_6)+\frac{3}{2}u_z(\widehat{g}_8+7\widehat{g}_9),\label{eq:collisionkernelg12}
\end{eqnarray}
\begin{eqnarray}
\widehat{g}_{13}&=&\frac{\omega_{13}}{8}\left[-(\widehat{\kappa}_{xyy}^{'}-\widehat{\kappa}_{xzz}^{'})+
2(u_y\widehat{\kappa}_{xy}^{'}-u_z\widehat{\kappa}_{xz}^{'})+
u_x(\widehat{\kappa}_{yy}^{'}-\widehat{\kappa}_{zz}^{'})\right.\nonumber\\
&&\left.-2\rho u_x(u_y^2-u_z^2)\right]+(u_y\widehat{g}_4-u_z\widehat{g}_5)+\frac{3}{4}u_x(-\widehat{g}_7+3\widehat{g}_8),\label{eq:collisionkernelg13}
\end{eqnarray}
\begin{eqnarray}
\widehat{g}_{14}&=&\frac{\omega_{14}}{8}\left[-(\widehat{\kappa}_{yzz}^{'}-\widehat{\kappa}_{xxy}^{'})+
2(u_z\widehat{\kappa}_{yz}^{'}-u_x\widehat{\kappa}_{xy}^{'})+
u_y(\widehat{\kappa}_{zz}^{'}-\widehat{\kappa}_{xx}^{'})\right.\nonumber\\
&&\left.-2\rho u_y(u_z^2-u_x^2)\right]+(-u_x\widehat{g}_4+u_z\widehat{g}_6)+\frac{3}{4}u_y(-\widehat{g}_7-3\widehat{g}_8),\label{eq:collisionkernelg14}
\end{eqnarray}
\begin{eqnarray}
\widehat{g}_{15}&=&\frac{\omega_{15}}{8}\left[-(\widehat{\kappa}_{xxz}^{'}-\widehat{\kappa}_{yyz}^{'})+
2(u_x\widehat{\kappa}_{xz}^{'}-u_y\widehat{\kappa}_{yz}^{'})+
u_z(\widehat{\kappa}_{xx}^{'}-\widehat{\kappa}_{yy}^{'})\right.\nonumber\\
&&\left.-2\rho u_z(u_x^2-u_y^2)\right.+(u_x\widehat{g}_5-u_y\widehat{g}_6)+\frac{3}{2}u_z \widehat{g}_7,\label{eq:collisionkernelg15}
\end{eqnarray}
\begin{eqnarray}
\widehat{g}_{16}&=&\frac{\omega_{16}}{12}\left[-(\widehat{\kappa}_{xxyy}^{'}+\widehat{\kappa}_{xxzz}^{'}+\widehat{\kappa}_{yyzz}^{'})+2u_x(\widehat{\kappa}_{xyy}^{'}+\widehat{\kappa}_{xzz}^{'})
+2u_y(\widehat{\kappa}_{xxy}^{'}\right.\nonumber\\
&&\left.+\widehat{\kappa}_{yzz}^{'})+
2u_z(\widehat{\kappa}_{xxz}^{'}+\widehat{\kappa}_{yyz}^{'})-u_x^2(\widehat{\kappa}_{yy}^{'}+\widehat{\kappa}_{zz}^{'}))
        -u_y^2(\widehat{\kappa}_{zz}^{'}+\widehat{\kappa}_{xx}^{'})\right.\nonumber\\
&&\left.
        -u_z^2(\widehat{\kappa}_{xx}^{'}+\widehat{\kappa}_{yy}^{'})
        -4(u_xu_y\widehat{\kappa}_{xy}^{'}+u_xu_z\widehat{\kappa}_{xz}^{'}+u_yu_z\widehat{\kappa}_{yz}^{'})\right.\nonumber\\
&&\left.+3\rho(u_x^2u_y^2+u_x^2u_z^2+u_y^2u_z^2)+
(\widetilde{\widehat{\kappa}}_{xx}\widetilde{\widehat{\kappa}}_{yy}+\widetilde{\widehat{\kappa}}_{xx}\widetilde{\widehat{\kappa}}_{zz}+
\widetilde{\widehat{\kappa}}_{yy}\widetilde{\widehat{\kappa}}_{zz})\right]\nonumber\\
&&
-\frac{4}{3}u_xu_y\widehat{g}_4-\frac{4}{3}u_xu_z\widehat{g}_5-\frac{4}{3}u_yu_z\widehat{g}_6+\frac{1}{2}(u_x^2-u_y^2)\widehat{g}_7+\frac{1}{2}(u_x^2+u_y^2-2u_z^2)\widehat{g}_8\nonumber\\
&&+(-7(u_x^2+u_y^2+u_z^2)-8)\widehat{g}_9+\frac{4}{3}u_x\widehat{g}_{10}+\frac{4}{3}u_y\widehat{g}_{11}+\frac{4}{3}u_z\widehat{g}_{12},\label{eq:collisionkernelg16}
\end{eqnarray}
\begin{eqnarray}
\widehat{g}_{17}&=&\frac{\omega_{17}}{24}\left[-(\widehat{\kappa}_{xxyy}^{'}+\widehat{\kappa}_{xxzz}^{'}-2\widehat{\kappa}_{yyzz}^{'})
+2\left(u_x\widehat{\kappa}_{xyy}^{'}+u_x\widehat{\kappa}_{xzz}^{'}+
       u_y\widehat{\kappa}_{xxy}^{'}+u_z\widehat{\kappa}_{xxz}^{'})-\right.\right.\nonumber\\
        &&\left.\left.2(u_y\widehat{\kappa}_{yzz}^{'}+u_z\widehat{\kappa}_{yyz}^{'})\right)
        -u_x^2(\widehat{\kappa}_{yy}^{'}+\widehat{\kappa}_{zz}^{'})
        -u_y^2(\widehat{\kappa}_{xx}^{'}-2\widehat{\kappa}_{zz}^{'})
        -u_z^2(\widehat{\kappa}_{xx}^{'}-2\widehat{\kappa}_{yy}^{'})\right.\nonumber\\
&&\left.-4(u_xu_y\widehat{\kappa}_{xy}^{'}+u_xu_z\widehat{\kappa}_{xz}^{'}-2u_yu_z\widehat{\kappa}_{yz}^{'})+
(\widetilde{\widehat{\kappa}}_{xx}\widetilde{\widehat{\kappa}}_{yy}+\widetilde{\widehat{\kappa}}_{xx}\widetilde{\widehat{\kappa}}_{zz}-2
\widetilde{\widehat{\kappa}}_{yy}\widetilde{\widehat{\kappa}}_{zz})\right.\nonumber\\
&&\left.+3\rho(u_x^2u_y^2+u_x^2u_z^2-2u_y^2u_z^2)\right]
-\frac{2}{3}u_xu_y\widehat{g}_4-\frac{2}{3}u_xu_z\widehat{g}_5+\frac{4}{3}u_yu_z\widehat{g}_6+\frac{1}{4}(u_x^2\nonumber\\
&&-u_y^2-3u_z^2-2)\widehat{g}_7+\frac{1}{4}(u_x^2-5u_y^2+u_z^2-2)\widehat{g}_8
+\frac{7}{4}(-2u_x^2+u_y^2+u_z^2)\widehat{g}_9\nonumber\\
&&+\frac{2}{3}u_x\widehat{g}_{10}-\frac{1}{3}u_y\widehat{g}_{11}-\frac{1}{3}u_z\widehat{g}_{12}-u_y\widehat{g}_{14}+u_z\widehat{g}_{15},\label{eq:collisionkernelg17}
\end{eqnarray}
\begin{eqnarray}
\widehat{g}_{18}&=&\frac{\omega_{18}}{8}\left[-(\widehat{\kappa}_{xxyy}^{'}-\widehat{\kappa}_{xxzz}^{'})
+2\left(u_x\widehat{\kappa}_{xyy}^{'}-u_x\widehat{\kappa}_{xzz}^{'}+
       u_y\widehat{\kappa}_{xxy}^{'}-u_z\widehat{\kappa}_{xxz}^{'})\right.\right.\nonumber\\
        &&\left.
        -u_x^2(\widehat{\kappa}_{yy}^{'}-\widehat{\kappa}_{zz}^{'})
        -u_y^2\widehat{\kappa}_{xx}^{'}
        +u_z^2\widehat{\kappa}_{xx}^{'}-4(u_xu_y\widehat{\kappa}_{xy}^{'}-u_xu_z\widehat{\kappa}_{xz}^{'})\right.\nonumber\\
&&+
(\widetilde{\widehat{\kappa}}_{xx}\widetilde{\widehat{\kappa}}_{yy}-\widetilde{\widehat{\kappa}}_{xx}\widetilde{\widehat{\kappa}}_{zz})\left.+3\rho(u_x^2u_y^2-u_x^2u_z^2)\right]
-2u_xu_y\widehat{g}_4+2u_xu_z\widehat{g}_5\nonumber\\
&&+\frac{3}{4}(u_x^2-u_y^2+u_z^2+\frac{2}{3})\widehat{g}_7+\frac{3}{4}(-3u_x^2-u_y^2+u_z^2-2)\widehat{g}_8
+\frac{21}{4}(-u_y^2+u_z^2)\widehat{g}_9\nonumber\\
&&+u_y\widehat{g}_{11}-u_z\widehat{g}_{12}+2u_x\widehat{g}_{13}-u_y\widehat{g}_{14}-u_z\widehat{g}_{15}.\label{eq:collisionkernelg18}
\end{eqnarray}

In the above, $\omega_4$, $\omega_5 \dots$ and $\omega_{18}$ are relaxation parameters of the central moments of different orders. Similar to the 2D central moment LBE~\cite{premnath2009incorporating}, we can apply the Chapman-Enskog expansion~\cite{chapman1990mathematical} to the above 3D cascaded LBE to show that it represents the Navier-Stokes equations. Some of the relaxation parameters in the collision model can be related to the transport coefficients of the momentum transfer. For example, those corresponding to the second-order moments control the shear viscosity $\nu$ and those related to the trace of the diagonal components (isotropic part) of such moments determine the bulk viscosity $\zeta$ of the fluid. That is, $\nu=c_s^2\left(\frac{1}{\omega^\nu}-\frac{1}{2}\right)$ where $\omega^{\nu}=\omega_{j}$ where $j=4,5,6,7,8$ and $\zeta=\frac{2}{3}c_s^2\left(\frac{1}{\omega^\zeta}-\frac{1}{2}\right)$ where $\omega^{\zeta}=\omega_{9}$. The rest of the parameters can be set either to $1$ (i.e., equilibration) or adjusted independently to carefully control and improve numerical stability by means of a linear stability analysis, while all satisfying the usual bounds $0<\omega_{\beta}<2$. In this work, for simplicity, we set the relaxation parameters for higher order moments to unity, i.e.,  $\omega_{j} =1$, where $j= 9, 10, 11,...,18$.

\subsection{\label{sec:cascadedcollisionexpressions}Nonlinear constitutive relations via local variations of relaxation times and local computation of strain rate tensor}
Then, for the simulation of non-Newtonian flows of power law fluids using the above 3D cascaded LBM, where the nonlinear constitutive relation prescribe the apparent viscosity dependent on the shear rate $|\dot{\gamma}_{ij}|$ , i.e., $\mu(|\dot{\gamma}_{ij}|)=\rho \nu(|\dot{\gamma}_{ij}|)$ (see Sec.~\ref{governing}), the relaxation times of the second order moments $\omega_4, \omega_5, \omega_6,  \omega_7$ and $\omega_8$ are adjusted locally as follows:
\begin{equation}
\omega^\nu(\bm{x},t)=\omega_4= \omega_5= \omega_6= \omega_7= \omega_8= \left[\frac{\mu(|\dot{\gamma}_{ij}|)}{\rho  c_s^2}+ \frac{1}{2}\right]^{-1},\label{eq:localvarationsrelaxationtimes}
\end{equation}
where $\mu(|\dot{\gamma}_{ij}|)$ can be obtained from Eq.~(\ref{eq6}), which is parametrized by the consistency coefficient $\mu_p$  and the power law index $n$. The magnitude of the shear rate $|\dot{\gamma}_{ij}|$ is related to the second invariant of the strain rate tensor $S_{ij}$ via Eq.~(\ref{eq:shearrateexpression}). The strain rate tensor components $S_{xx},S_{yy},S_{zz},S_{xy},S_{yz}$ and $S_{xz}$ can be obtained locally using non-equilibrium moments as follows.

A Chapman-Enskog analysis~\cite{chapman1990mathematical} of the 3D cascaded LBE, whose details are omitted here for brevity but follows as an extension of those presented in~\cite{premnath2009incorporating,hajabdollahi2018galilean}, relates the components of the second order non-equilibrium moments to the components of the symmetric part of the velocity gradient tensor. The results can be summarized as follows:
\begin{eqnarray*}
\widehat{m}^{(1)}_4 &\approx& -\frac{1}{3\omega_4}\rho(\partial_yu_x+\partial_xu_y),\quad \widehat{m}^{(1)}_5 \approx -\frac{1}{3\omega_5}\rho(\partial_zu_x+\partial_xu_z),\\
\widehat{m}^{(1)}_6 &\approx& -\frac{1}{3\omega_6}\rho(\partial_zu_y+\partial_yu_z), \quad \widehat{m}^{(1)}_7 \approx -\frac{2}{3\omega_7}\rho(\partial_xu_x-\partial_yu_y),\\
\widehat{m}^{(1)}_8 &\approx& -\frac{2}{3\omega_8}\rho(\partial_xu_x-\partial_zu_z), \quad \widehat{m}^{(1)}_9 \approx -\frac{2}{3\omega_9}\rho(\partial_xu_x+\partial_yu_y+\partial_zu_z),
\end{eqnarray*}
where the non-equilibrium moments $\widehat{m}^{(1)}_4, \widehat{m}^{(1)}_5,\ldots,\widehat{m}^{(1)}_9$ are given by
\begin{subequations}
\begin{eqnarray}
\widehat{m}^{(1)}_4 &=& \widehat{\kappa}^{'}_{xy}-\rho u_xu_y,\\
\widehat{m}^{(1)}_5 &=& \widehat{\kappa}^{'}_{xz}-\rho u_xu_z,\\
\widehat{m}^{(1)}_6 &=& \widehat{\kappa}^{'}_{yz}-\rho u_yu_z,\\
\widehat{m}^{(1)}_7 &=& (\widehat{\kappa}^{'}_{xx}-\widehat{\kappa}^{'}_{yy})-\rho (u_x^2-u_y^2),\\
\widehat{m}^{(1)}_8 &=& (\widehat{\kappa}^{'}_{xx}-\widehat{\kappa}^{'}_{zz})-\rho (u_x^2-u_z^2),\\
\widehat{m}^{(1)}_9 &=& (\widehat{\kappa}^{'}_{xx}+\widehat{\kappa}^{'}_{yy}+\widehat{\kappa}^{'}_{zz})-(\rho (u_x^2+u_y^2+u_z^2)+\rho).
\end{eqnarray}
\end{subequations}
From the above, and considering Eq.~(\ref{eq:localvarationsrelaxationtimes}) and invoking the definition of the strain rate tensor $S_{ij}=\frac{1}{2}(\partial_ju_i+\partial_iu_j)$, its components can be solved in terms of the non-equilibrium moments $\widehat{m}^{(1)}_4, \widehat{m}^{(1)}_5,\ldots,\widehat{m}^{(1)}_9$ as
\begin{subequations}
\begin{eqnarray}
S_{xy} &=& -\frac{3}{2\rho}\omega^\nu(\bm{x},t)\widehat{m}^{(1)}_4,\\
S_{xz} &=& -\frac{3}{2\rho}\omega^\nu(\bm{x},t)\widehat{m}^{(1)}_5,\\
S_{yz} &=& -\frac{3}{2\rho}\omega^\nu(\bm{x},t)\widehat{m}^{(1)}_6,\\
S_{xx} &=& -\frac{1}{2\rho}\left\{\omega^\nu(\bm{x},t)\left[\widehat{m}^{(1)}_7+\widehat{m}^{(1)}_8\right]+\omega^\zeta\widehat{m}^{(1)}_9\right\},\\
S_{yy} &=& -\frac{1}{2\rho}\left\{\omega^\nu(\bm{x},t)\left[-2\widehat{m}^{(1)}_7+\widehat{m}^{(1)}_8\right]+\omega^\zeta\widehat{m}^{(1)}_9\right\},\\
S_{zz} &=& -\frac{1}{2\rho}\left\{\omega^\nu(\bm{x},t)\left[\widehat{m}^{(1)}_7-2\widehat{m}^{(1)}_8\right]+\omega^\zeta\widehat{m}^{(1)}_9\right\}.
\end{eqnarray}\label{eq:strainratetensor}
\end{subequations}
The above two sets of equations enable local computation of the strain rate tensor in simulating non-Newtonian flows and they are naturally suitable for parallel computing, as opposed to using finite difference approximations.

\subsection{\label{sec:postcollisiondistributionfunctions}Post-collision distribution functions}
Finally, evaluating the elements of $(\tensor{K}\cdot \mathbf{\widehat{g}})_{\alpha}$ in the collision step of the 3D cascaded LBM (see Eqs. (\ref{eq:cascadecollision1}) and (\ref{eq:cascadedcollision1})), the post collision distribution functions can be obtained. These are summarized as follows:
\begin{eqnarray}
\widetilde{f}_{0}&=&f_{0}+\left[\widehat{g}_0-30\widehat{g}_9+12\widehat{g}_{16}\right], \nonumber\\
\widetilde{f}_{1}&=&f_{1}+\left[\widehat{g}_0+\widehat{g}_1+\widehat{g}_7+\widehat{g}_8-11\widehat{g}_9
-4\widehat{g}_{10}-4\widehat{g}_{16}-4\widehat{g}_{17}\right], \nonumber\\
\widetilde{f}_{2}&=&f_{2}+\left[\widehat{g}_0-\widehat{g}_1+\widehat{g}_7+\widehat{g}_8-11\widehat{g}_9
+4\widehat{g}_{10}-4\widehat{g}_{16}-4\widehat{g}_{17}\right], \nonumber\\
\widetilde{f}_{3}&=&f_{3}+\left[\widehat{g}_0+\widehat{g}_2-\widehat{g}_7+\widehat{g}_8-11\widehat{g}_9
-4\widehat{g}_{11}-4\widehat{g}_{16}+2\widehat{g}_{17}-2\widehat{g}_{18}\right], \nonumber\\
\widetilde{f}_{4}&=&f_{4}+\left[\widehat{g}_0-\widehat{g}_2-\widehat{g}_7+\widehat{g}_8-11\widehat{g}_9
+4\widehat{g}_{11}-4\widehat{g}_{16}+2\widehat{g}_{17}-2\widehat{g}_{18}\right], \nonumber\\
\widetilde{f}_{5}&=&f_{5}+\left[\widehat{g}_0+\widehat{g}_3-2\widehat{g}_8-11\widehat{g}_9-4\widehat{g}_{12}
-4\widehat{g}_{16}+2\widehat{g}_{17}+2\widehat{g}_{18}\right], \nonumber\\
\widetilde{f}_{6}&=&f_{6}+\left[\widehat{g}_0-\widehat{g}_3-2\widehat{g}_8-11\widehat{g}_9+4\widehat{g}_{12}
-4\widehat{g}_{16}+2\widehat{g}_{17}+2\widehat{g}_{18}\right], \nonumber\\
\widetilde{f}_{7}&=&f_{7}+\left[\widehat{g}_0+\widehat{g}_1+\widehat{g}_2+\widehat{g}_4+2\widehat{g}_8+8\widehat{g}_9
+\widehat{g}_{10}+\widehat{g}_{11}+\widehat{g}_{13}-\widehat{g}_{14}+\widehat{g}_{16}+\widehat{g}_{17}+\widehat{g}_{18}\right],\nonumber\\
\widetilde{f}_{8}&=&f_{8}+\left[\widehat{g}_0-\widehat{g}_1+\widehat{g}_2-\widehat{g}_4+2\widehat{g}_8+8\widehat{g}_9
-\widehat{g}_{10}+\widehat{g}_{11}-\widehat{g}_{13}-\widehat{g}_{14}+\widehat{g}_{16}+\widehat{g}_{17}+\widehat{g}_{18}\right],\nonumber\\
\widetilde{f}_{9}&=&f_{9}+\left[\widehat{g}_0+\widehat{g}_1-\widehat{g}_2-\widehat{g}_4+2\widehat{g}_8+8\widehat{g}_9
+\widehat{g}_{10}-\widehat{g}_{11}+\widehat{g}_{13}+\widehat{g}_{14}+\widehat{g}_{16}+\widehat{g}_{17}+\widehat{g}_{18}\right],\nonumber\\
\widetilde{f}_{10}&=&f_{10}+\left[\widehat{g}_0-\widehat{g}_1-\widehat{g}_2+\widehat{g}_4+2\widehat{g}_8+8\widehat{g}_9
-\widehat{g}_{10}-\widehat{g}_{11}-\widehat{g}_{13}+\widehat{g}_{14}+\widehat{g}_{16}+\widehat{g}_{17}+\widehat{g}_{18}\right],\nonumber\\
\widetilde{f}_{11}&=&f_{11}+\left[\widehat{g}_0+\widehat{g}_1+\widehat{g}_3+\widehat{g}_5+\widehat{g}_7
-\widehat{g}_8+8\widehat{g}_9+\widehat{g}_{10}+\widehat{g}_{12}-\widehat{g}_{13}+\widehat{g}_{15}+\widehat{g}_{16}+\widehat{g}_{17}-\widehat{g}_{18}\right],\nonumber\\
\widetilde{f}_{12}&=&f_{12}+\left[\widehat{g}_0-\widehat{g}_1+\widehat{g}_3-\widehat{g}_5+\widehat{g}_7
-\widehat{g}_8+8\widehat{g}_9-\widehat{g}_{10}+\widehat{g}_{12}+\widehat{g}_{13}+\widehat{g}_{15}+\widehat{g}_{16}+\widehat{g}_{17}-\widehat{g}_{18}\right],\nonumber\\
\widetilde{f}_{13}&=&f_{13}+\left[\widehat{g}_0+\widehat{g}_1-\widehat{g}_3-\widehat{g}_5+\widehat{g}_7
-\widehat{g}_8+8\widehat{g}_9+\widehat{g}_{10}-\widehat{g}_{12}-\widehat{g}_{13}-\widehat{g}_{15}+\widehat{g}_{16}+\widehat{g}_{17}-\widehat{g}_{18}\right],\nonumber\\
\widetilde{f}_{14}&=&f_{14}+\left[\widehat{g}_0-\widehat{g}_1-\widehat{g}_3+\widehat{g}_5+\widehat{g}_7
-\widehat{g}_8+8\widehat{g}_9-\widehat{g}_{10}-\widehat{g}_{12}+\widehat{g}_{13}-\widehat{g}_{15}+\widehat{g}_{16}+\widehat{g}_{17}-\widehat{g}_{18}\right],\nonumber\\
\widetilde{f}_{15}&=&f_{15}+\left[\widehat{g}_0+\widehat{g}_2+\widehat{g}_3-\widehat{g}_6-\widehat{g}_7
-\widehat{g}_8+8\widehat{g}_9+\widehat{g}_{11}+\widehat{g}_{12}+\widehat{g}_{14}-\widehat{g}_{15}+\widehat{g}_{16}-2\widehat{g}_{17}\right],\nonumber\\
\widetilde{f}_{16}&=&f_{16}+\left[\widehat{g}_0-\widehat{g}_2+\widehat{g}_3-\widehat{g}_6-\widehat{g}_7
-\widehat{g}_8+8\widehat{g}_9-\widehat{g}_{11}+\widehat{g}_{12}-\widehat{g}_{14}-\widehat{g}_{15}+\widehat{g}_{16}-2\widehat{g}_{17}\right],\nonumber\\
\widetilde{f}_{17}&=&f_{17}+\left[\widehat{g}_0+\widehat{g}_2-\widehat{g}_3-\widehat{g}_6-\widehat{g}_7
-\widehat{g}_8+8\widehat{g}_9+\widehat{g}_{11}-\widehat{g}_{12}+\widehat{g}_{14}+\widehat{g}_{15}+\widehat{g}_{16}-2\widehat{g}_{17}\right],\nonumber\\
\widetilde{f}_{18}&=&f_{18}+\left[\widehat{g}_0-\widehat{g}_2-\widehat{g}_3+\widehat{g}_6-\widehat{g}_7
-\widehat{g}_8+8\widehat{g}_9-\widehat{g}_{11}-\widehat{g}_{12}-\widehat{g}_{14}+\widehat{g}_{15}+\widehat{g}_{16}-2\widehat{g}_{17}\right].\nonumber
\end{eqnarray}
This is then followed by the streaming step via Eq.~(\ref{eq:cascadedstreaming1}), after which the hydrodynamic fields of the 3D non-Newtonian fluid flow can be obtained from Eqs.~(\ref{eq:hydrodynamicfields1}) and (\ref{eq:hydrodynamicfields2}).

\subsection{\label{sec:implementationstrategy}Implementation strategy of the algorithm}
The streaming step (Eq.~(\ref{eq:cascadedstreaming1})) is a perfect shift operation of the distribution functions involving data transfer (in memory) to adjacent lattice nodes along the particle velocity directions. As such, this operation remains the same regardless of the collision model employed. The main computational effort involving floating point operations is related to the collision step (Eq.~(\ref{eq:cascadedcollision1}) via Eq.~(\ref{eq:cascadecollision1})). In particular, this latter step involves (i) the computation of the various raw moments (Eq.~(\ref{eq:rawmomentdistributionfunction1})), (ii) change of moments under cascaded collision via central moment relaxation (i.e., the components of $\mathbf{\widehat{g}}$ as given in Eqs.~(\ref{eq:collisionkernelg4})-(\ref{eq:collisionkernelg18})), and (iii) mapping the changes in moments to obtain the post-collision distribution function (via $(\tensor{K}\cdot \mathbf{\widehat{g}})_{\alpha}$ as given in the section above to compute $\widetilde{f}_{\alpha}$, where $\alpha=0,1,2,\ldots,18$). In this regard, various optimization strategies should be fully utilized in implementing each of these substeps.

For example, the computation of the raw moments as expressed by Eq.~(\ref{eq:rawmomentdistributionfunction1}) needs to be implemented by only involving the non-zero elements after grouping the common factors present in the basis vectors $\bra{\mathbf{e_{x}}^m\mathbf{e_{y}}^n\mathbf{e_{z}}^p}$ in performing the inner product with the distribution function $\ket{\mathbf{f}}$. In the computation of the changes in moments under collision $\mathbf{\widehat{g}}$, the common subexpressions are evaluated once and then stored as separate scalar variables and reused. Moreover, all common numerical constants appearing as coefficients involving multiplications and divisions are to be computed once and assigned to additional variables at the beginning of simulations. In this substep, for implementing the non-linear constitutive relationships for non-Newtonian fluids via the local variations of the relaxation times, the required strain rate tensor $S_{ij}$ is obtained locally as shown in Sec.~\ref{sec:cascadedcollisionexpressions} by reusing the already available second order non-equilibrium moments, which are needed in calculating $\widehat{g}_4$ through $\widehat{g}_9$. Then, in order to map back the moment changes to the distribution functions as represented by $(\tensor{K}\cdot \mathbf{\widehat{g}})_{\alpha}$, no direct matrix-vector multiplication should be performed. Rather, the nature of the elements of the mapping matrix $\tensor{K}$ involving integers, especially the presence of many zeroes and various common factors, should be fully exploited. In other words, only the relevant subexpressions with non-zero elements, by grouping terms with common factors, needs to be computed and reused when possible. When these optimization strategies are employed, as discussed later in Sec.~\ref{sec:Comparison}, it only incurs a moderate additional computational overhead when compared to the common SRT-LB formulation, while delivering significant improvements in numerical stability.

\section{\label{section_numerical_tests}Results and discussion}
To validate our 3D cascaded LBM for non-Newtonian flows, we consider the following benchmark problems for which either the analytical solutions and/or numerical benchmark results available: (i) non-Newtonian Poiseuille flow of a power law fluid, (ii) three-dimensional fully developed flow in a square duct, and (iii) power-law fluid flow in a cubic lid-driven cavity (see Fig.~\ref{fig:img_2}). For the first two problems, there exist analytical solutions, while for the third problem, we use the recent benchmark numerical data of~\cite{jin2017gpu} for comparison.

\subsection{Poiseuille flow of power-law fluid}
\begin{figure}[htbp]
\centering
\advance\leftskip-2.0cm
    \subfloat{
        \includegraphics[scale=.8]{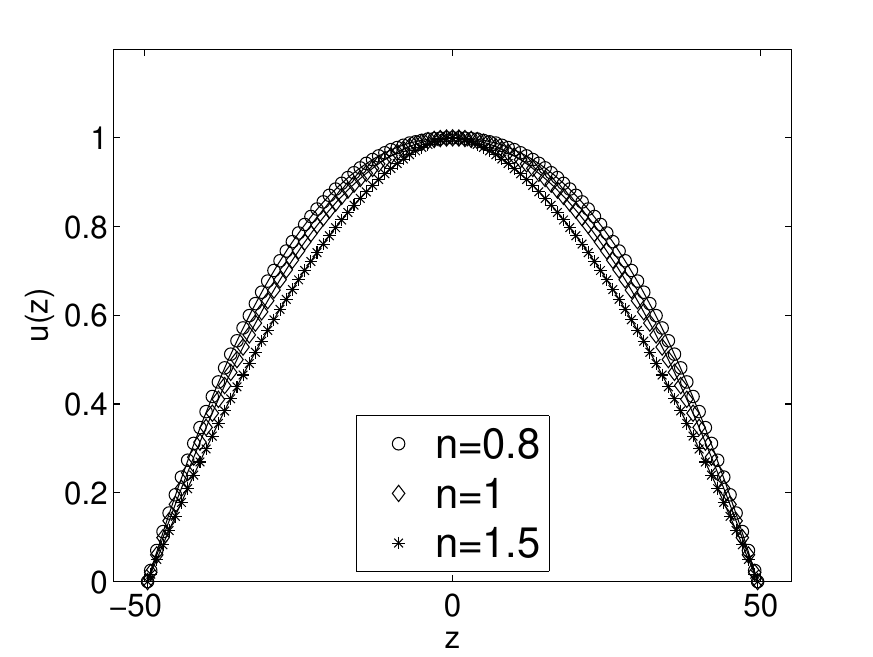}
        } \\
    \caption{Comparison of the normalized velocity profiles of power law fluids in a channel computed using 3D Cascaded LBM  against the analytical solution for power law index $ n=0.8,1.0,$ and $1.5$ at $\mbox{Re} = 100$.}
    \label{fig:img_3}
\end{figure}

First, we consider the flow of power-law fluids between two parallel plates separated by a height $H$ and driven by a constant body force. Periodic boundary conditions along the streamwise direction and spanwise directions and a half-way bounce-back boundary condition is applied on the fixed walls. The flow is driven by a body force of $F_x=1.0\times 10^{-6}$, which  represents the pressure gradient $-\frac{\partial P}{\partial x}$. The characteristic dimensionless Reynolds number $\mbox{Re}$ for this problem may be written as $\mbox{Re}=\rho(\frac{H}{2})^n u_{max}^{2-n}/\mu_p$, where $u_{max}$ is the maximum velocity of the flow occurring midway between the two plates (see below). We consider a grid resolution of $3\times 3\times 101$ for resolving the domain of the flow of power law fluids with $\mbox{Re}=100$ and $n=0.8, 1.0$ and $1.5$, thereby encompassing both shear thinning and shear thickening fluids. The analytical solution for this non-Newtonian flow problem is given by
\begin{equation*}
u(z)=\frac{n}{n+1} \left(- \frac{1}{\mu_p}  \frac{\partial P}{\partial x} \right)^{1/n}\left[ \left(\frac{H}{2}\right)^{(n+1)/n}-\left|z\right|^{(n+1)/n} \right],
\end{equation*}
where the maximum fluid velocity is obtained using
\begin{equation*}
u_{max}=\frac{n}{n+1}\left(-\frac{1}{\mu_p} \frac{\partial P}{\partial x}\right)^
{1/n}\left(\frac{H}{2} \right)^{(n+1)/n}.
\end{equation*}
Figure~\ref{fig:img_3} presents a comparison between the computed velocity profiles for different values of the power law index obtained using the 3D cascaded LBM implemented on a D3Q19 lattice with the analytical solution. It is clear that the numerical simulation results for different power law fluids are in very good agreement with the analytical solution.

\begin{figure}[h!]
\centering
\advance\leftskip-2.0cm
    \subfloat{
        \includegraphics[scale=.7] {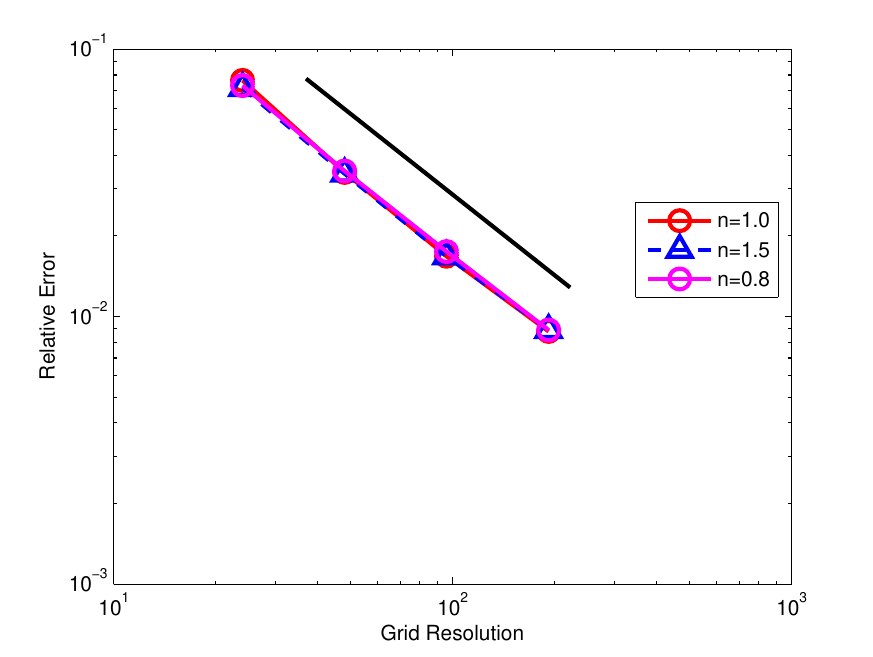}
        \label{fig:img4L} } \\
    \caption{Relative global error against the variation in the grid spacing, plotted in log-log scales, for the simulation of power law fluid flow in a channel at $n=0.8$, $n=1.0$ and $1.5$ obtained using the 3D cascaded LBM. The solid line (black) represents an ideal slope of $-2.0$.}
    \label{fig:img3_4}
\end{figure}

In addition, Fig.~\ref{fig:img3_4} demonstrates that our 3D cascaded LBM is second order accurate for the simulation of power law fluids. In this grid convergence test, we have used diffusive scaling to establish asymptotic convergence of our scheme to the incompressible macroscopic non-Newtonian fluid flow equations of shear thinning, Newtonian and shear thickening fluids.

\subsection{Three-dimensional fully developed flow in a square duct}
\begin{figure}[htbp]
\centering
\advance\leftskip-2.0cm
    \subfloat[]{
        \includegraphics[width=.65\textwidth] {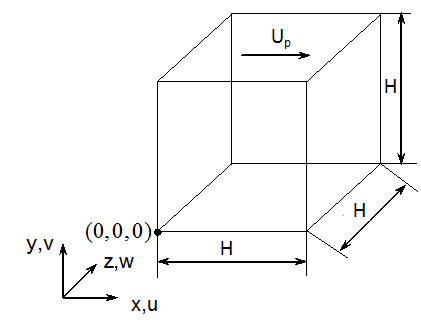}
        \label{fig:img_2a} } \hspace*{-1em}

     \subfloat[]{
        \includegraphics[width=.65\textwidth] {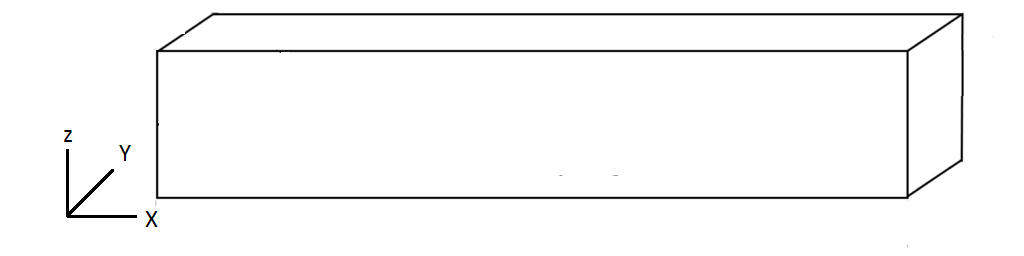}
        \label{fig:img_2b} } \\
        \advance\leftskip0cm
    \caption{Schematics of the non-Newtonian benchmark flow problems. (a) Power law fluid flow in a lid-driven cubic cavity, (b) 3D fully developed flow in a square duct. }
    \label{fig:img_2}
\end{figure}

\begin{figure}[htbp]
\centering
\advance\leftskip-1.7cm
    \subfloat []{
        \includegraphics[scale=0.7, angle=0] {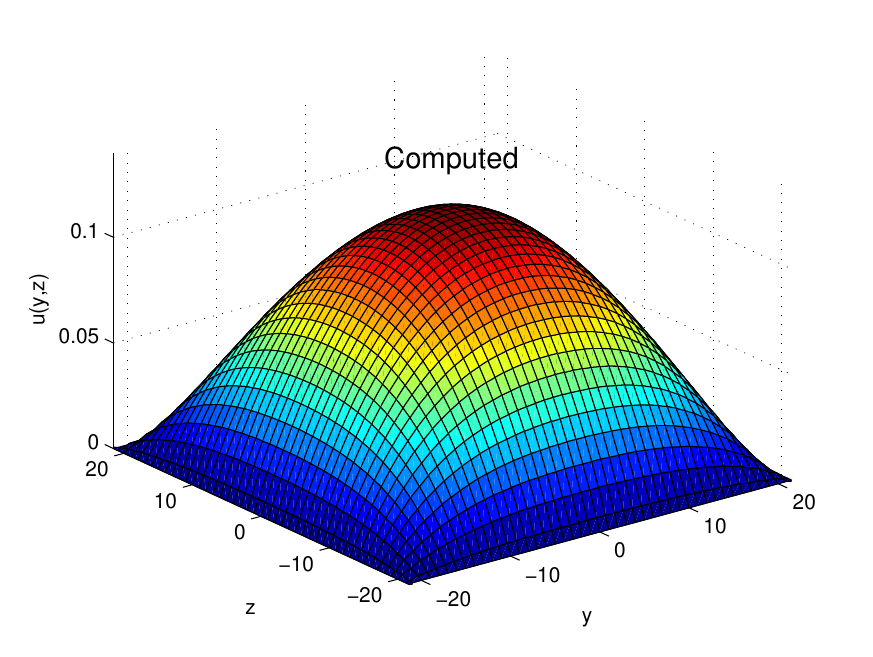}
        \label{fig:img_4a} } \\
    \subfloat[]{
        \includegraphics[scale=0.7, angle=0] {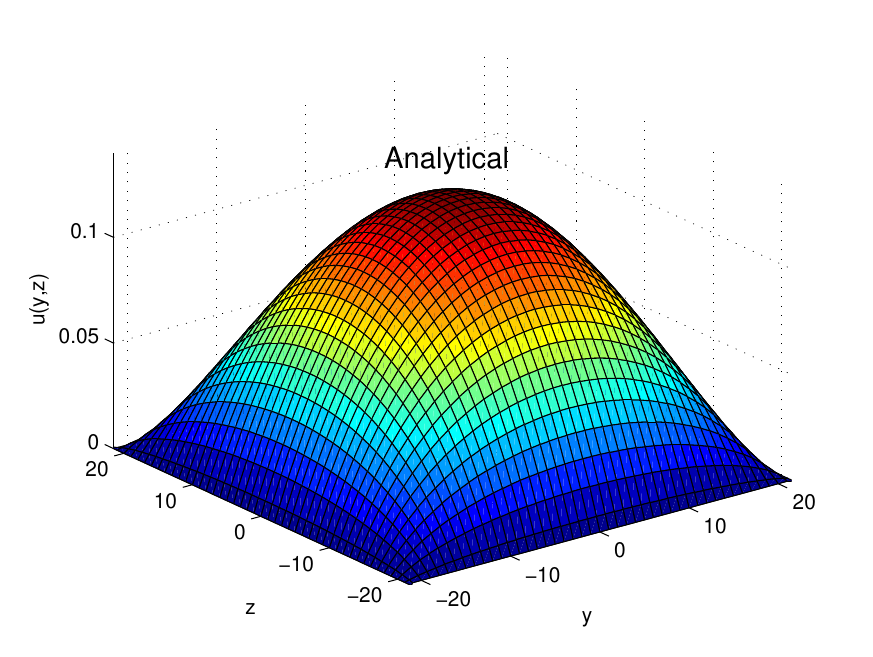}
        \label{fig:img_4b} } \\
    \caption{{  Flow through a square duct with side length 2a subjected to a constant body force: Comparison of surface contours of the velocity field for Reynolds number Re = 80 (a) computed by the D3Q19 formulation of the cascaded LBM with forcing term with (b) analytical solution.}}
    \label{fig:img_4}
\end{figure}

\begin{figure}[htbp]
\centering
\advance\leftskip-2.0cm
    \subfloat{
            \includegraphics[scale=0.7, angle=0]{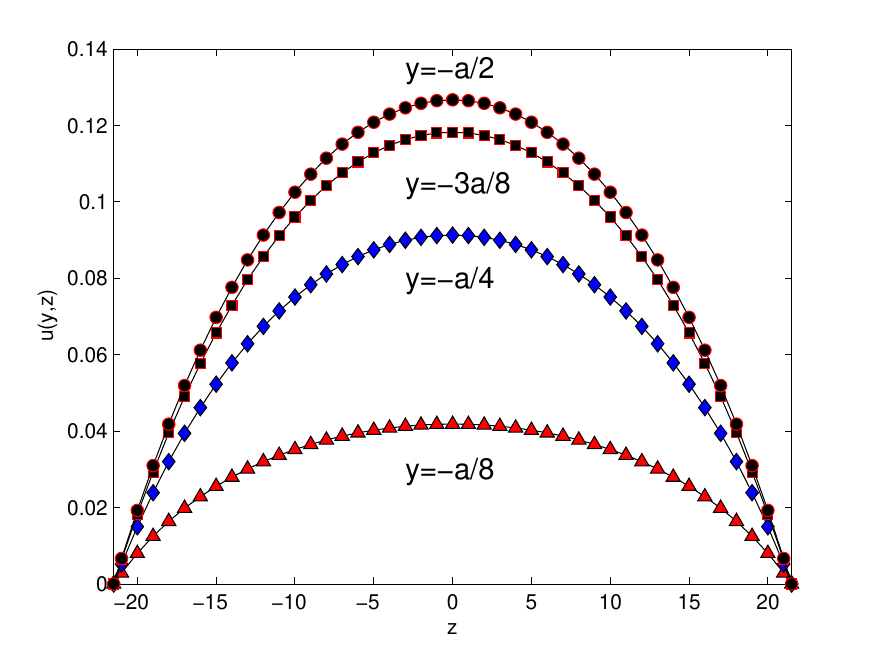} 
         }
    \caption{Flow through a square duct with side length 2a subjected to a constant body force: Comparison of velocity profiles computed by the D3Q19 formulation of the cascaded LBM with forcing term (symbols) with analytical solution (lines) at different locations in the duct cross-section for Reynolds
number Re = 80.}
    \label{fig:img_5a}
    \end{figure}

\begin{figure}[htbp]
\centering
\advance\leftskip-2.0cm
    \subfloat{
        \includegraphics[scale=0.7, angle=0] {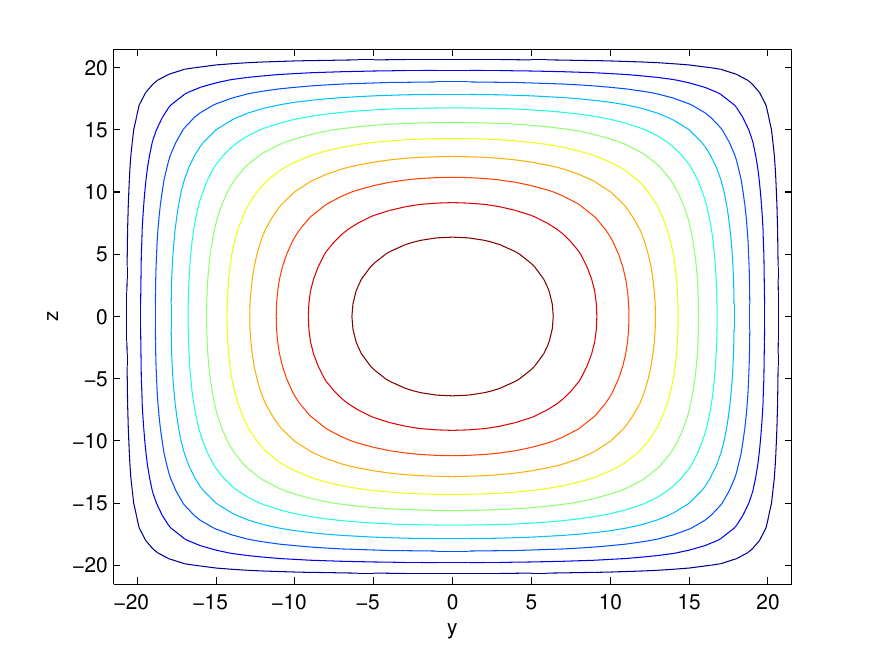}
        \label{fig:img_5b} } \\
    \caption{{  Iso-speed contours of the duct velocity field for Reynolds number Re = 80.}}
    \label{fig:img_5}
\end{figure}

\begin{figure}[htbp]
\centering
\advance\leftskip-2.0cm
    \subfloat{
            \includegraphics[scale=0.7, angle=0]{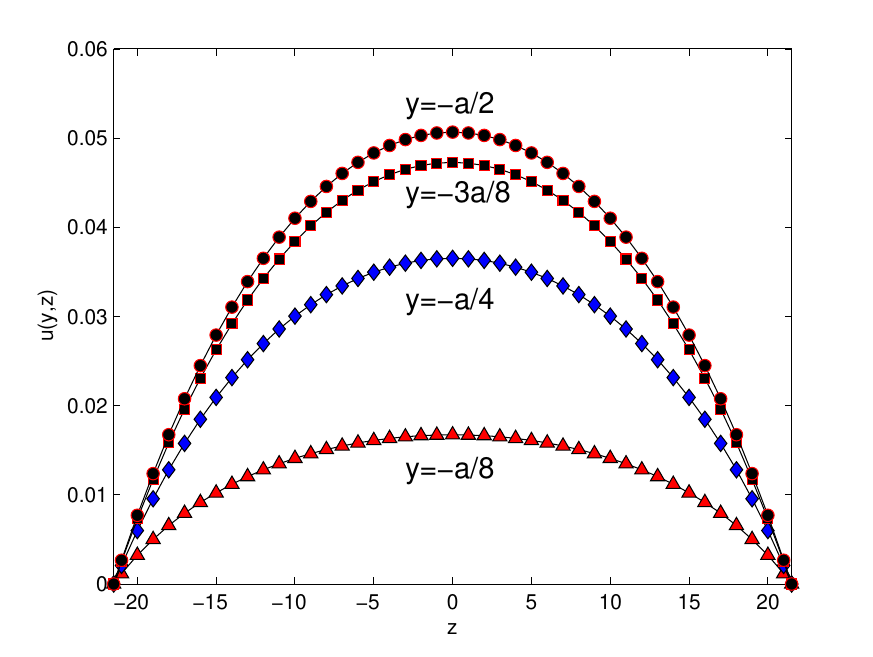} 
         }
    \caption{Flow through a square duct with side length 2a subjected to a constant body force: Comparison of velocity profiles computed by the D3Q19 formulation of the cascaded LBM with forcing term (symbols) with analytical solution (lines) at different locations in the duct cross-section for Reynolds number Re = 20.}
    \label{fig:img_7a}
    \end{figure}

\begin{figure}[htbp]
\centering
\advance\leftskip-2.0cm
    \subfloat{
        \includegraphics[scale=0.7, angle=0] {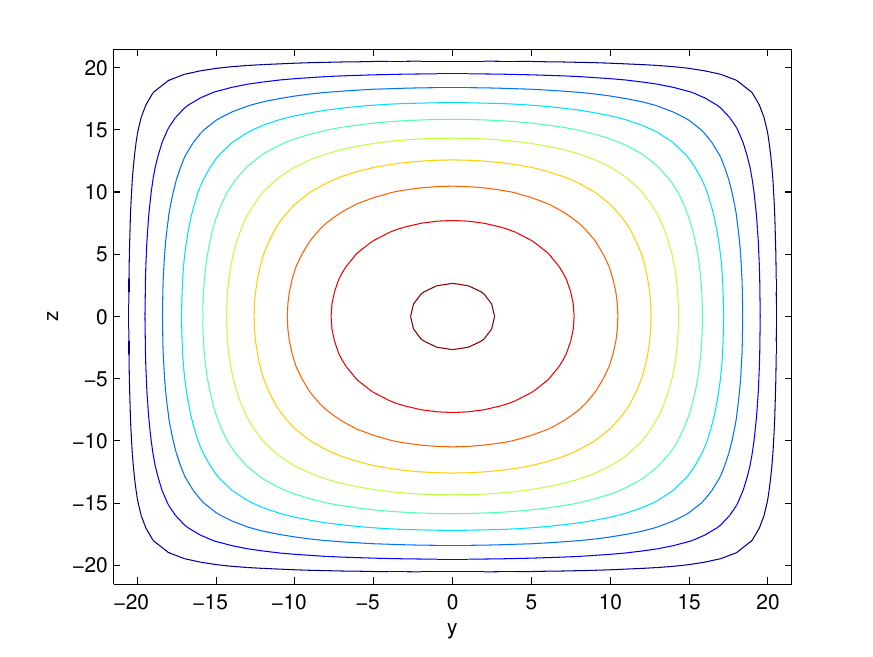}
        \label{fig:img_7b} } \\
    \caption{{  Iso-speed contours of the duct velocity field for Reynolds number Re = 20. }}
    \label{fig:img_7}
\end{figure}

The fluid flow through a 3D square duct (see Fig.~\ref{fig:img_2b}) is considered as the next example. While an analytical solution for such a problem does not exist for power law fluid at all values of the power law index, an exact solution for the Newtonian fluid flow case ($n=1$) exists (see below), which could be used to test our 3D cascaded LB formulation on a D3Q19 lattice. The cross section of the square duct is defined by $-a \leq y \leq a$ and $-a\leq z \leq a$, where $a$ is the half width of the duct, and $x$ is the streamwise flow direction. Periodic boundary conditions are applied at the inlet and outlet, whereas a no-slip boundary condition is adopted on the wall boundaries by using the standard half-way bounce back approach. The Reynolds number $\mbox{Re}$ for this problem is defined based on the maximum fluid velocity and the duct half width. The flow is driven by applying a body force of $F_x=1.0\times 10^{-6}$, and a grid resolution of $3\times 45 \times 45$ is employed to represent the flow domain. The body force is implementation is discussed in Ref.~\cite{hajabdollahi2018symmetrized}, who is adopted for the D3Q19 lattice. The analytical solution for the velocity field of the fully developed square duct flow can be expressed in terms of the following infinite series based on harmonic functions~\cite{white2006viscous}:
\begin{equation*}\label{eqn12}
u(y,z)=\frac{16a^2F_x}{\rho \nu \pi^3}\sum_{n=1}^{\infty}(-1)^{(n-1)}\left[1-\frac{\cosh\left(\frac{(2n-1)\pi z}{2a}\right)}{\cosh\left(\frac{(2n-1)\pi}{2}\right)}\right]\frac{\cos\left(\frac{(2n-1)\pi y}{2a}\right)}{(2n-1)^3}.
\end{equation*}
Figures~\ref{fig:img_4a} and~\ref{fig:img_4b} show the comparison of the surface contours of the velocity field computed using the 3D cascaded LBM using a D3Q19 lattice for a Reynolds number of 80 and the analytical solution (see the equation above). It is clear that the computed results are in good agreement with the exact solution, which shows a paraboloid distribution for the velocity profile. In order to get a better perspective for comparison, Figs.~\ref{fig:img_5a} and~\ref{fig:img_7a} show the comparisons of the computed velocity profiles at different locations in the duct for $\mbox{Re}=80$ and $20$, respectively, which show excellent agreement of our 3D cascaded LBM with the analytical solution. In addition, the iso-speed contours of the velocity field at these two $\mbox{Re}$ are presented in Figs.~\ref{fig:img_5} and~\ref{fig:img_7}, respectively, which are consistent with expectations.


\begin{figure}[htbp]
\centering
\advance\leftskip-1.7cm
    \subfloat [Re=100]{
        \includegraphics[width=.55\textwidth] {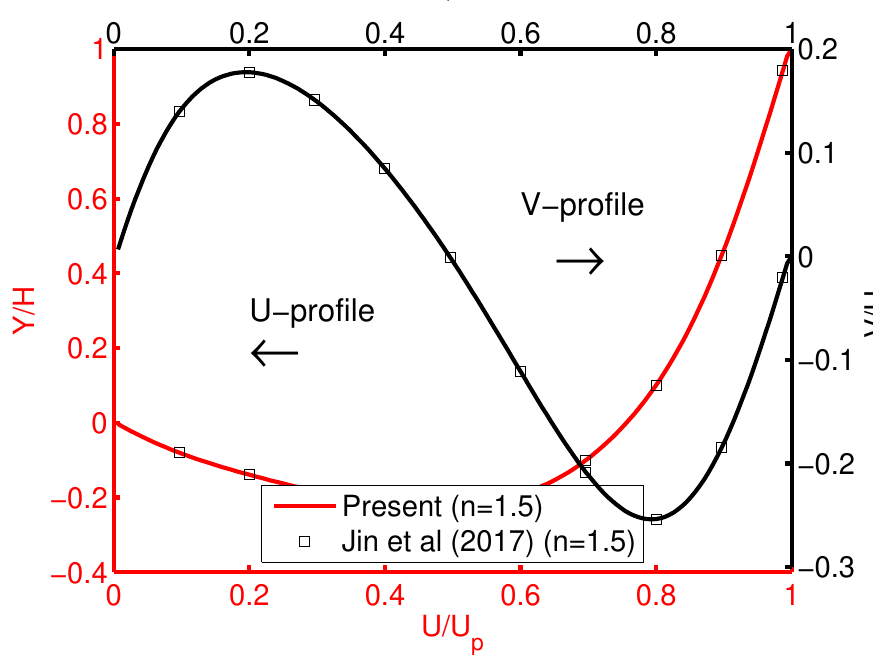}
        \label{fig:img_10a} } 
    \subfloat[Re=400]{
        \includegraphics[width=.55\textwidth] {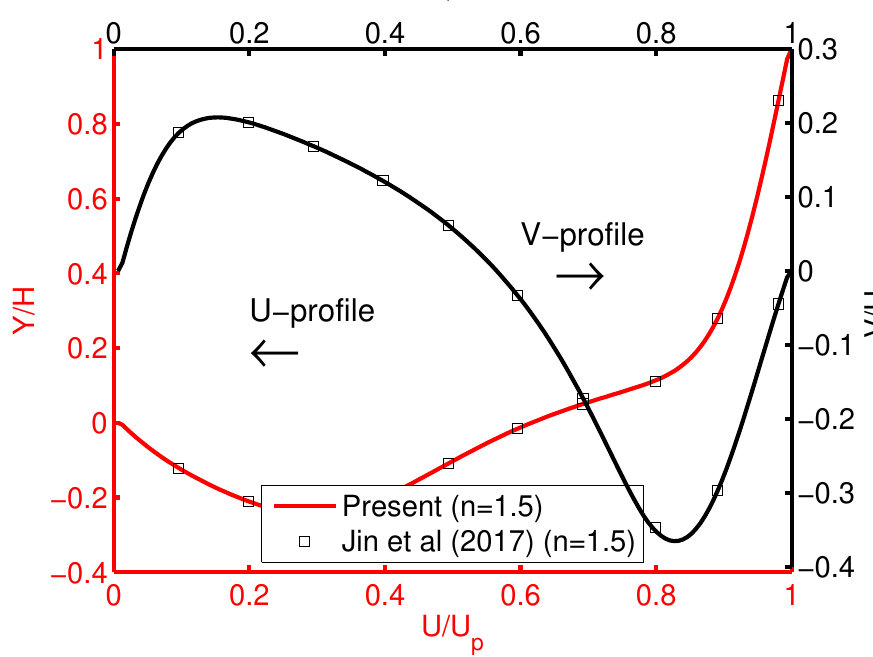}
        \label{fig:img_10b} } \\
                \advance\leftskip0cm
        \subfloat[Re=1000]{
        \includegraphics[width=.55\textwidth] {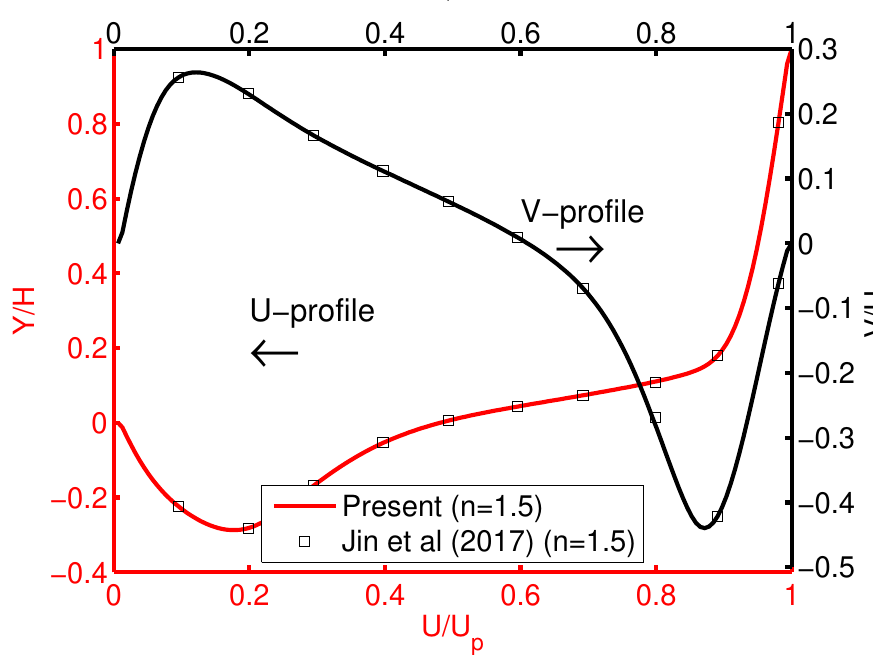}
        \label{fig:img_10c} } \\
    \caption{{  Computed velocity component $u/U_p$ along the vertical centerline and component $v/U_p$ along the horizontal centerline obtained using the cascaded LB method (lines) and compared with the benchmark solution of Jin et al (2017) (symbols) for lid-driven cubic cavity flow of power law fluids. Reynolds number $Re=100,400,1000$ and the power law index $n=1.5$.}}
    \label{fig:img_10}
\end{figure}
\begin{figure}[htbp]
\centering
\advance\leftskip-1.7cm
    \subfloat [Re=100]{
        \includegraphics[width=.55\textwidth] {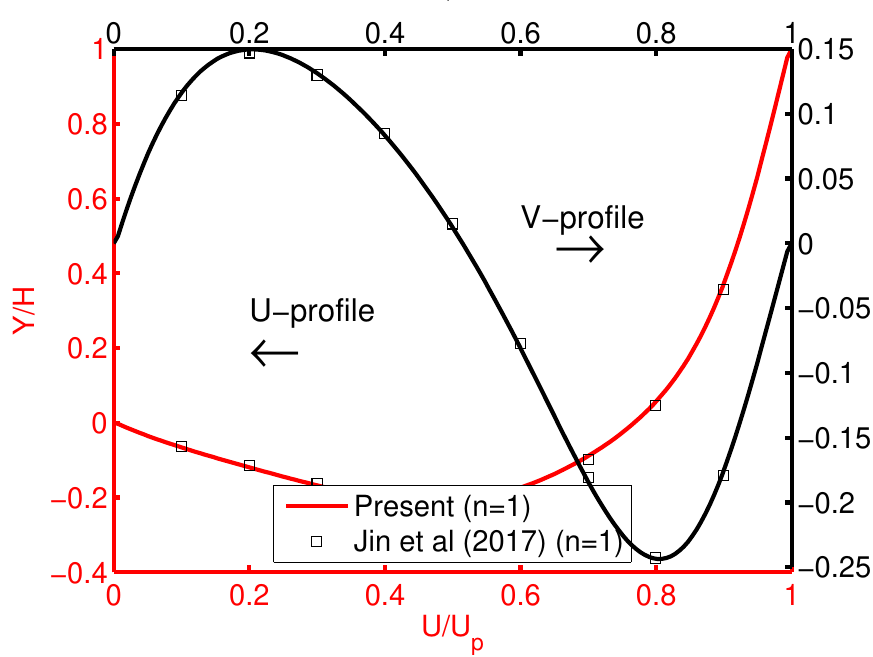}
        \label{fig:img_11a} } 
    \subfloat[Re=400]{
        \includegraphics[width=.55\textwidth] {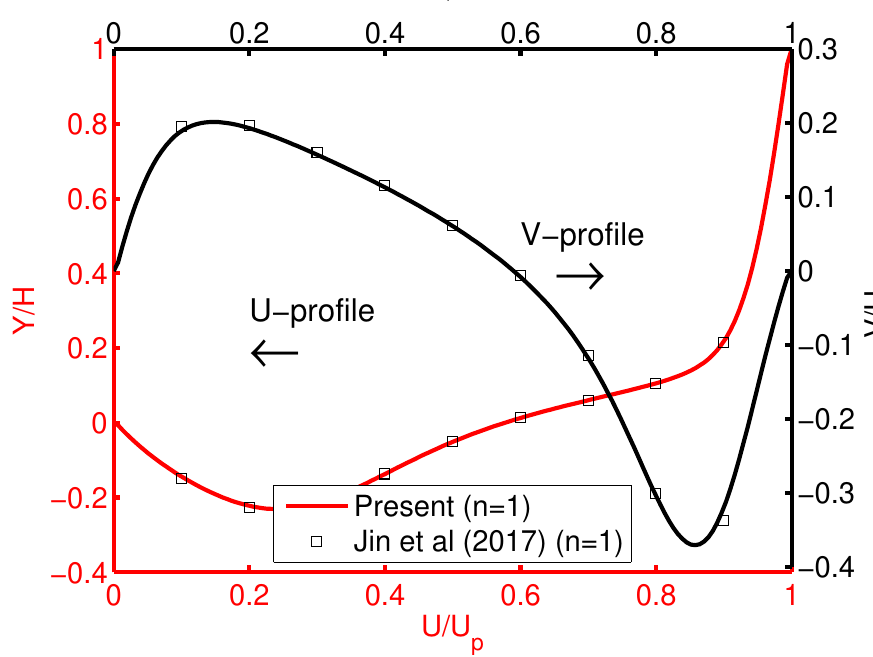}
        \label{fig:img_11b} } \\
                \advance\leftskip0cm
        \subfloat[Re=1000]{
        \includegraphics[width=.55\textwidth] {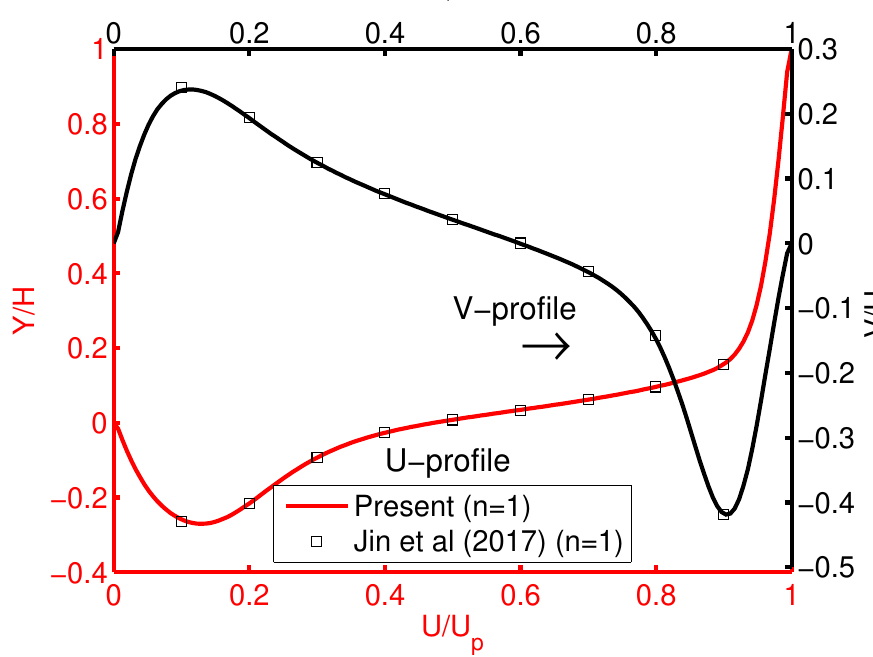}
        \label{fig:img_11c} } \\
    \caption{{   Computed velocity component $u/U_p$ along the vertical centerline and component $v/U_p$ along the horizontal centerline obtained using the cascaded LB method (lines) and compared with the benchmark solution of Jin et al (2017) (symbols) for lid-driven cubic cavity flow of power law fluids. Reynolds number $Re=100,400,1000$ and the power law index $n=1.0$.}}
    \label{fig:img_11}
\end{figure}
\begin{figure}[htbp]
\centering
\advance\leftskip-1.7cm
    \subfloat [Re=100]{
        \includegraphics[width=.55\textwidth] {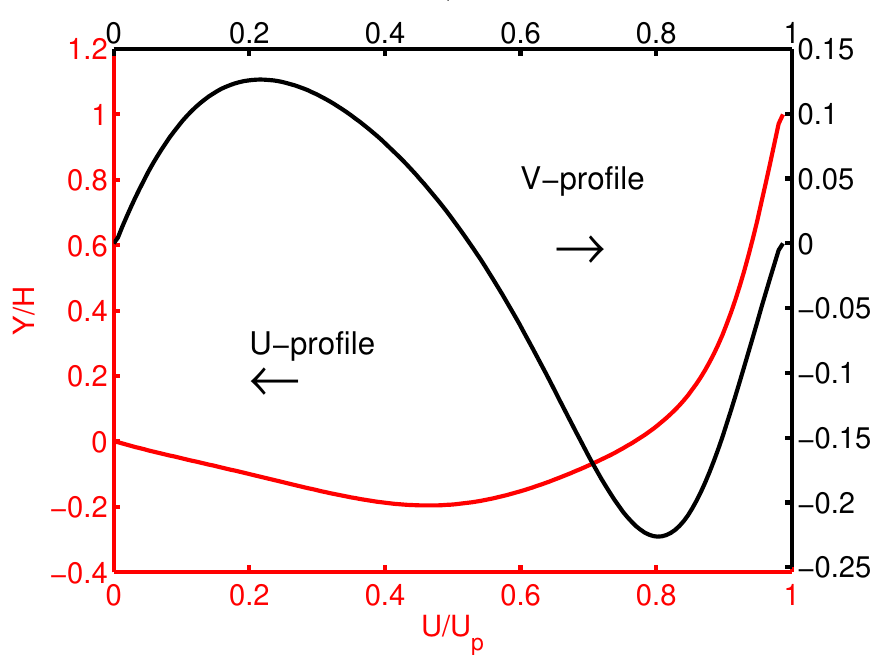}
        \label{fig:img_12a} } 
    \subfloat[Re=400]{
        \includegraphics[width=.55\textwidth] {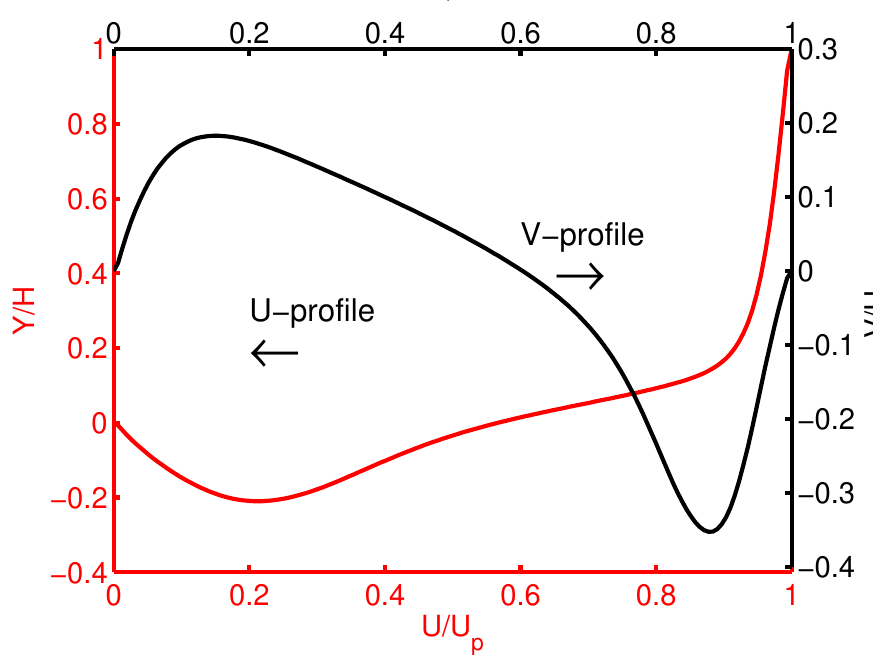}
        \label{fig:img_12b} } \\
                \advance\leftskip0cm
        \subfloat[Re=1000]{
        \includegraphics[width=.55\textwidth] {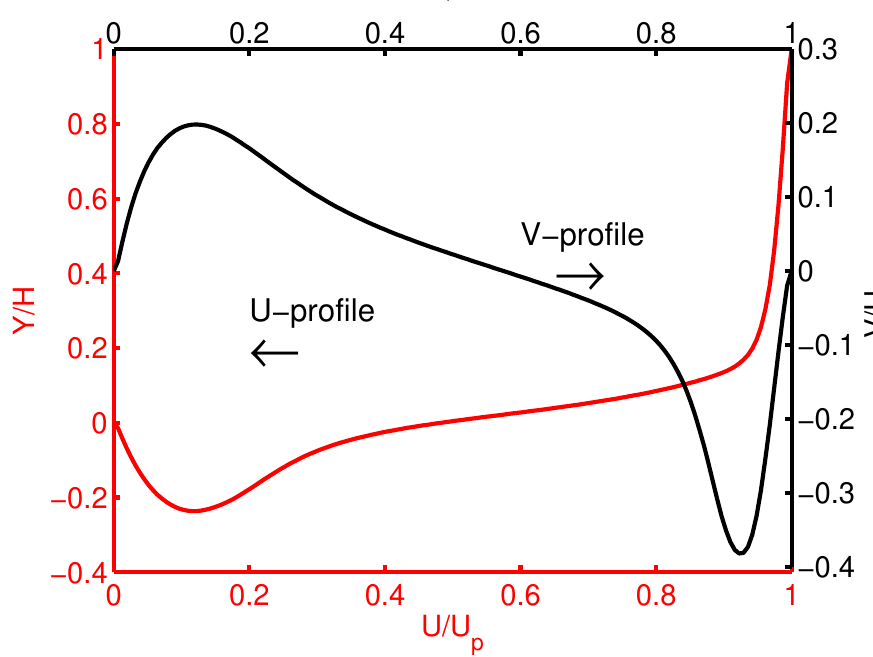}
        \label{fig:img_12c} } \\
    \caption{{  Computed velocity component $u/U_p$ along the vertical centerline and component $v/U_p$ along the horizontal centerline obtained using the cascaded LB method (lines) and compared with the benchmark solution of Jin et al (2017) (symbols) for lid-driven cubic cavity flow of power law fluids. Reynolds number $Re=100,400,1000$ and the power law index $n=0.8$.}}
    \label{fig:img_12}
\end{figure}

\subsection{3D Power-law fluid flow in a lid-driven cubic cavity}
The 3D lid-driven flow of non-Newtonian fluids in a cubic cavity is a classic benchmark problem with complex flow features and is a stringent test to validate numerical methods. The geometric configuration of this problem is shown in Fig.~\ref{fig:img_2a}, where a power-law fluid is enclosed within a cubic cavity of side length $H$. The upper surface of the cavity or the lid is subjected to a uniform velocity $U_p$, while all the remaining walls of the cavity are maintained stationary. As noted in the introduction, only very few studies exist on the 3D non-Newtonian flows of power law fluids in cubic cavities driven by its top lid. A recent work reported by~\cite{jin2017gpu} presented high quality results based on fine enough grid resolutions using a finite volume method, which will be used as a benchmark solution for comparison.

We employ a grid resolution of $128\times 128 \times 128 $ in our simulations, which was found to lead to grid independent results for the Reynolds number cases considered in this regard. The no-slip boundary conditions on the walls are imposed using the standard half-way bounce back scheme, with the motion of the lid represented by applying a momentum correction to the respective bounce back condition. The characteristic Reynolds number for this problem is defined by $\mbox{Re}=\rho H^n U_p^{2-n}/\mu_p$. We consider the simulation of cubic cavity flow at three different Reynolds numbers of $\mbox{Re}=100, 400$ and $1000$. In each case, we consider three types of non-Newtonian fluids with power law index $n=1.5, 1.0$ and $0.8$, thereby encompassing both shear thickening and shear thinning cases. In all these cases the flow was found to be stationary and hence all the simulations were run until the steady state was reached.

The numerical results for the computed profiles of the velocity components along the horizontal and vertical centerlines in the plane $z=0.5H$ at the above three $\mbox{Re}$ for the power law $n=1.5, 1.0$ and $0.8$ are shown in Figs.~\ref{fig:img_10},~\ref{fig:img_11} and ~\ref{fig:img_12}, respectively. While for the Newtonian fluid, the viscosity is a constant, the local viscosity varies appreciably depending on the strain rate tensor components that changes due to variations in the vortical flow structures inside the cubic cavity. As a result, in the case of shear thickening fluid, as the flow encounters high shear zones near the moving lid, its effective viscosity increases in these regions, diffusing its momentum (and conversely for the shear thinning case) and altering the velocity profiles in the respective cases. Hence, the boundary layer becomes thicker for fluid with $n>1$ when compared to the case where $n<1$. In addition, it is found that as the power law index $n$ increases, the peak magnitude of the velocity component $v$ along the horizontal centerline increases. These observations are consistent with the recent findings of~\cite{jin2017gpu}. Overall, the computed results obtained using the present 3D central moment LBM are in good quantitative agreement with the benchmark results of~\cite{jin2017gpu}.

\section{\label{sec:Comparison}Comparison of numerical stability and computational performance of different collision models}
We now make a direct comparison of the numerical stability of our cascaded LB scheme with other LB formulations for the simulation of 3D lid-driven cubic cavity flow simulations. Due to it being a shear driven flow and accompanied by geometric singularity around the corners, this flow configuration serves as a good benchmark problem to test the stability of the numerical schemes.

\begin{figure}[htbp]
\centering
\advance\leftskip-1.7cm
    \subfloat [$n=0.8$]{
        \includegraphics[width=.55\textwidth] {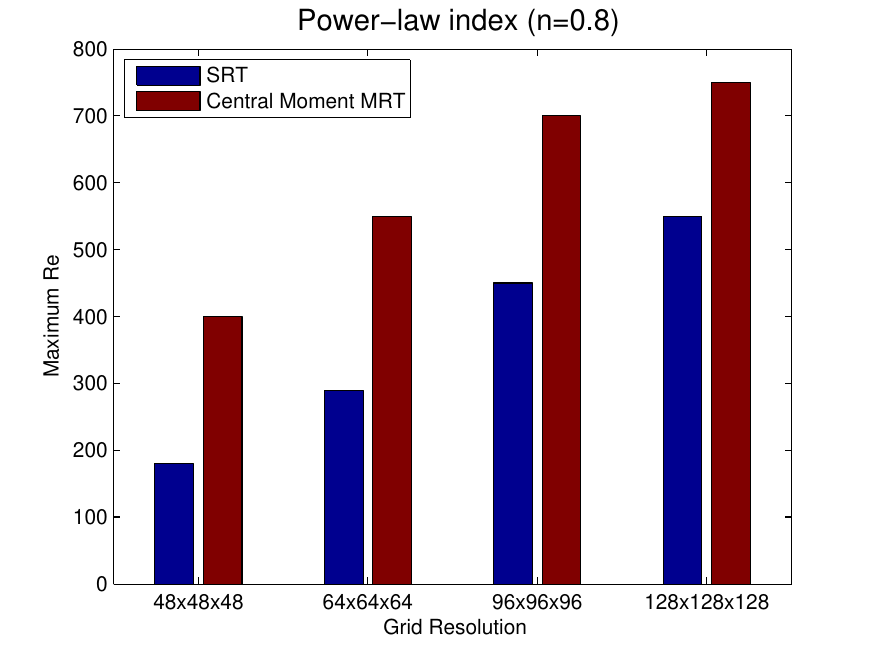}
        \label{fig:img_13a} } 
    \subfloat[$n=1.0$]{
        \includegraphics[width=.55\textwidth] {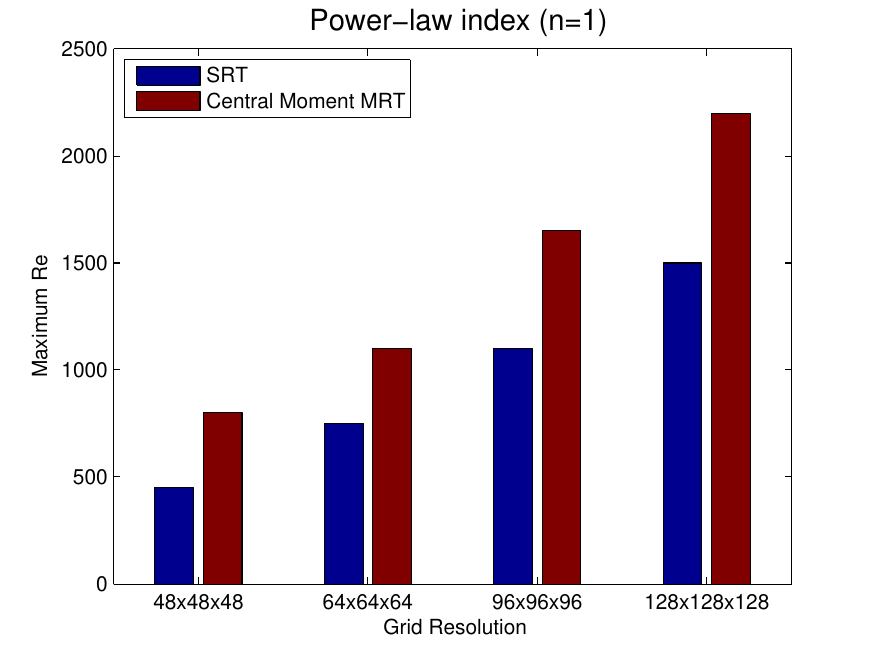}
        \label{fig:img_13b} } \\
                \advance\leftskip0cm
        \subfloat[$n=1.5$]{
        \includegraphics[width=.55\textwidth] {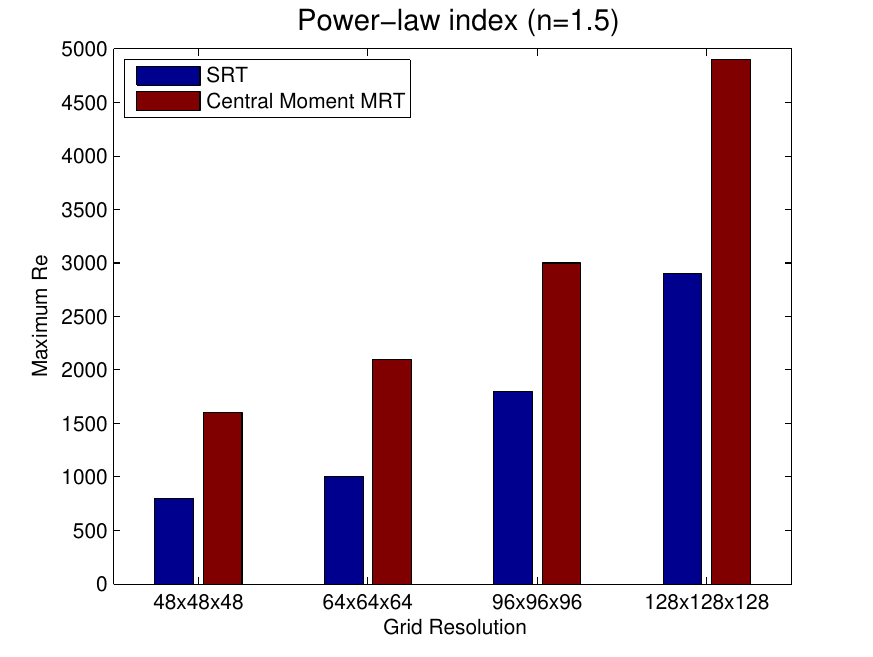}
        \label{fig:img_13c} } \\
    \caption{{Comparison of the maximum Reynolds number for numerical stability of the 3D SRT-LBM and the 3D cascaded LBM based on central moments for simulation of non-Newtonian lid driven cubic cavity flow with $n = 0.8, 1.0, 1.5$.}}
    \label{fig:img_13}
\end{figure}

In the first case study, we consider our approach and the popular SRT-LBM both implemented using the D3Q19 lattice, where for a chosen grid resolution and a fixed lid velocity, the relaxation parameter for the second order moments that controls viscosity was decreased gradually until the computation becomes unstable. It may be noted that in the case of cascaded LBM, as mentioned earlier, the relaxation parameters of higher kinetic moments are set to unity for simplicity. Then, the numerical instability is deemed to occur when the global error of the velocity field becomes exponentially large. Figure~\ref{fig:img_13} shows the maximum Reynolds number that could be attained before the computations become unstable for both the collision models. Results are presented for three different values of the power law index ($n=0.8, 1.0$ and $1.5$) encompassing both shear thinning and shear thickening fluids at various set of grid resolutions ($48^3$, $64^3$, $96^3$ and $128^3$). It is evident that the cascaded LBM computations achieve significantly higher Reynolds numbers than the SRT model for the same grid resolution and for a given type of fluid or a power law index, and thus is more stable. Here, a note regarding the associated computational cost is in order. When the implementation strategy involving various optimizations as discussed in Sec.~\ref{sec:implementationstrategy} are fully exploited, the additional computational overhead of using the cascaded LB formulation when compared to the SRT-LBM is moderate. In particular, it was found that the time required to update a node for the cascaded LBM is about $30\%-40\%$ more than that for the SRT-LBM, which is consistent with other related studies (e.g.,~\cite{Ning2015,hajabdollahi2019cascaded}). On the other hand, the cascaded LBM is significantly more stable. For example, when $n=1$, the SRT-LBM requires a finer grid with a resolution of $96^3$ to achieve a similar Reynolds number as that of the cascaded LBM, which utilizes a coarser grid with a resolution of $64^3$. In other words, this means that the SRT-LBM uses about 3.4 times greater number of grid nodes than the cascaded LBM to achieve similar Reynolds number while maintaining numerical stability. It is thus clear that the cascaded LBM delivers an overall superior computational performance when compared to the SRT-LBM, even for the special case where we have set their relaxation times for the higher order moments to unity. On the other hand, as demonstrated in a previous 2D study~\cite{Ning2015}, the stability characteristics of the cascaded LB formulation is enhanced further by tuning the relaxation time for the trace of the second order moments, which controls the bulk viscosity, and/or the relaxation times of the third and higher order kinetic moments. Hence, it is expected that further improvements in stability and overall computational performance are possible by optimizing the relaxation parameters of the higher kinetic moments of the 3D cascaded LBM for non-Newtonian flow simulations.

\begin{figure}[htbp]
\centering
\advance\leftskip-1.7cm
    \subfloat [$n=0.8$]{
        \includegraphics[width=.60\textwidth] {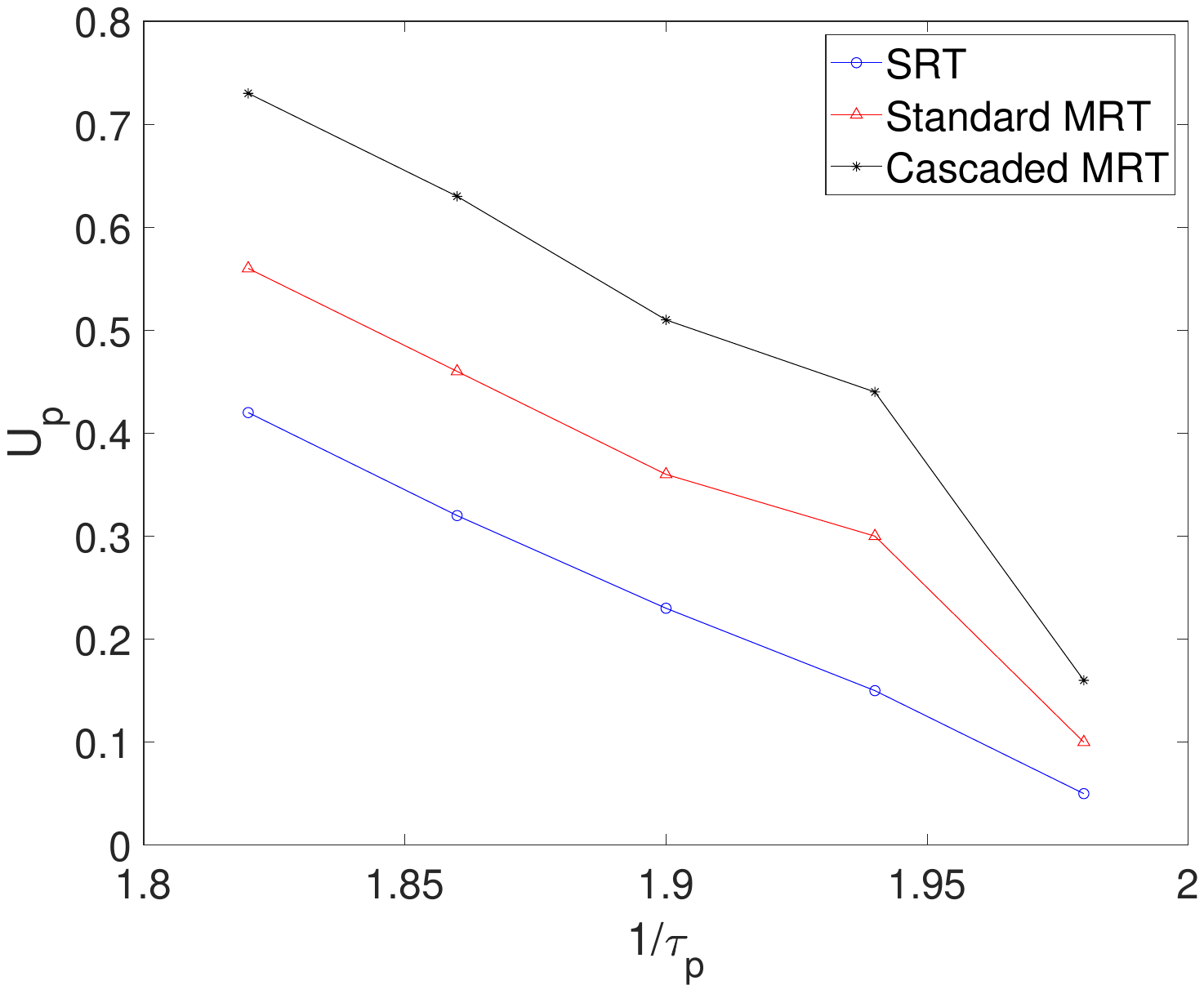}
        \label{fig:img_14a} } 
    \subfloat[$n=1.0$]{
        \includegraphics[width=.60\textwidth] {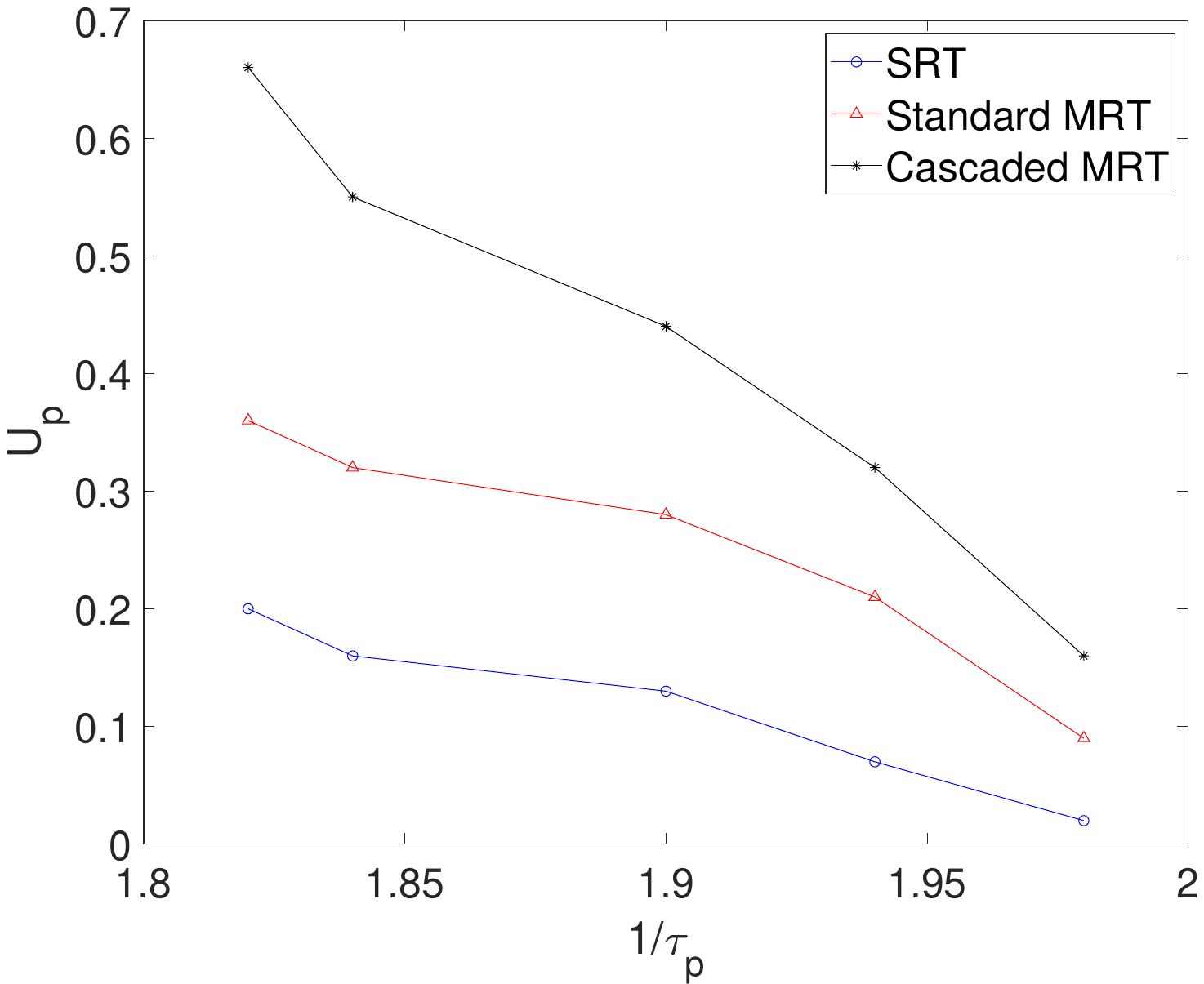}
        \label{fig:img_14b} } \\
                \advance\leftskip0cm
        \subfloat[$n=1.5$]{
        \includegraphics[width=.60\textwidth] {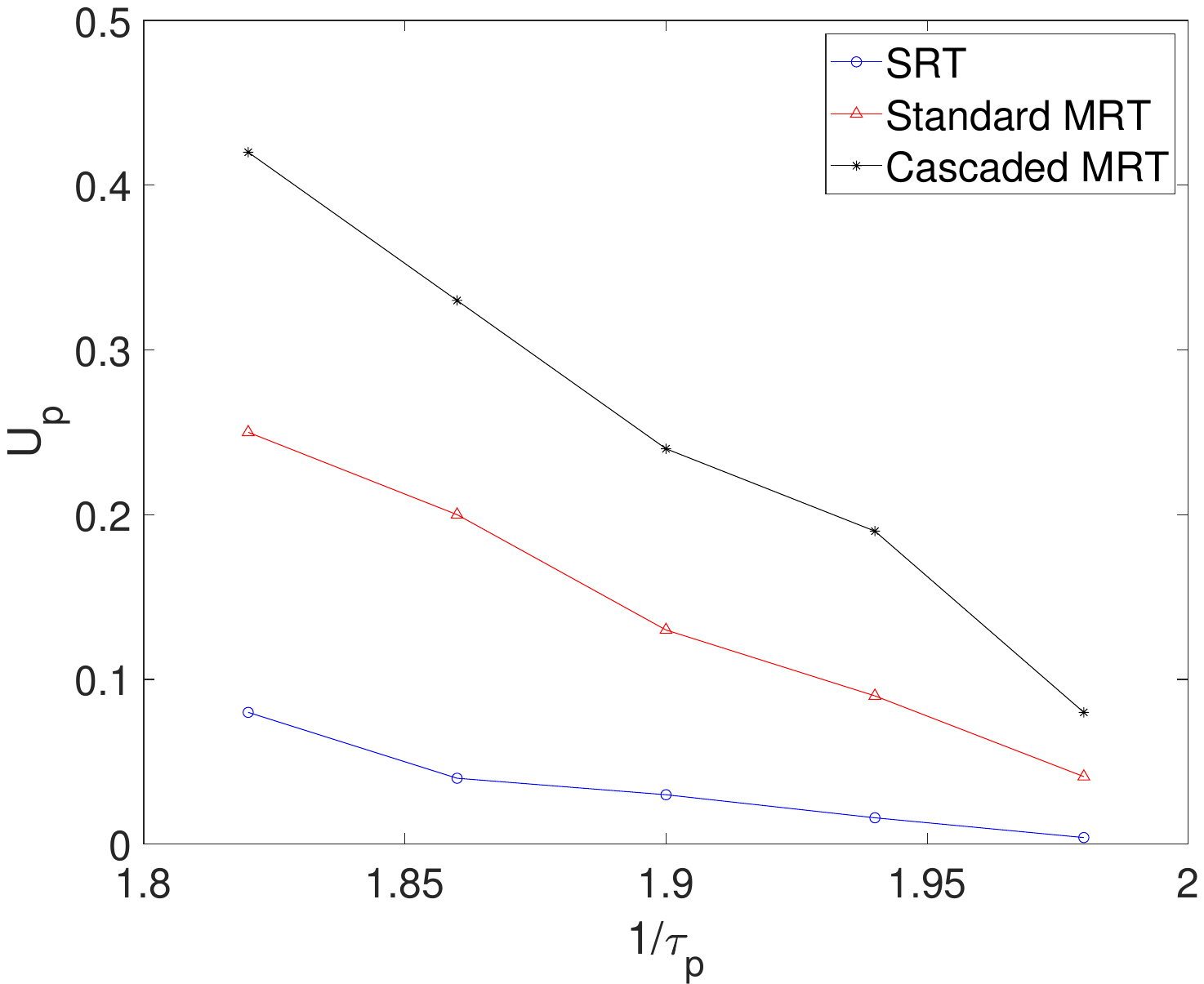}
        \label{fig:img_14c} } \\
    \caption{{Numerical stability test results showing the maximum threshold velocity of the top plate (lid) $U_p$ in a 3D cubic cavity flow of non-Newtonian fluids for a grid resolution of $64^3$ at different values of the consistency coefficient of the power law fluid $\mu_p$ defined in Eq.~(\ref{eq5}) (or equivalently $\tau_p$ via $\mu_p = c_s^2(\tau_p-1/2)$). Comparisons are made between the SRT-LBM, standard MRT-LBM based on raw moments and the present cascaded MRT-LBM based on central moments for different values of the power law index $n = 0.8, 1.0$ and $1.5$.}}
    \label{fig:img_14}
\end{figure}

We will now present a second case study involving a numerical stability test that determines the maximum threshold velocity of the lid at a chosen grid resolution using different LB formulations, viz., the SRT-LBM, standard MRT-LBM based on raw moments and the present cascaded MRT-LBM based on central moments, for simulation of the 3D cubic cavity flow at different values of the consistency coefficient $\mu_p$ and the index $n$ of the power law fluid (see Eq.~(\ref{eq5})). In this regard, a fixed grid resolution of $64^3$ is used and the consistency coefficient $\mu_p$ is chosen equivalently via specifying a parameter $\tau_p$ using $\mu_p = c_s^2(\tau_p-1/2)$, and for each LB scheme, following the strategy given in Refs. ~\cite{ luo2011numerics, Ning2015}, the maximum lid velocity $U_p$ that can maintain stable computations for 50,000 time steps is determined. In the SRT-LBM, all the distribution functions relax at the same rate; the standard MRT-LBM based on raw moments~\cite{ d2002multiple} is adapted for non-Newtonian flow simulations by locally adjusting the relaxation rates of the second order moments, while the relaxation parameters for the higher order kinetic moments are chosen based on the values given in Ref.~\cite{d2002multiple}; in the case of the cascaded MRT-LBM based on central moments, the relaxation rates for the second order moments vary locally as before, while those for the higher order moments are set to unity. Figure~\ref{fig:img_14} presents the comparisons of the threshold lid velocity obtained using the various LB schemes for the simulation of power law fluids with index $n$ equal to $0.8, 1.0$ and $1.5$ for this alternate stability test. In general, for all the values of the power law index, encompassing shear thinning and shear thickening fluids, the maximum lid velocities for stability achieved using the cascaded MRT-LBM are found to be considerably higher than those compared to the other collision models. Equivalently, this means that it can sustain stable simulations of non-Newtonian flows at significantly higher Reynolds numbers. This is consistent with the observation that the use of central moments in the cascaded LBM naturally preserves the Galilean invariance of the moments supported independently by the lattice and involving higher order fluid velocity terms in its equilibria when compared to the SRT or the standard MRT-LBM based on raw moments, whose equilibria are generally based on the fluid velocity terms up to the second order. As a result, the use of the 3D cascaded LBM presents significant advantages for simulations of non-Newtonian flows of power law fluids.

\section{\label{sec:summary}Summary and Conclusions}
Complex fluid flows satisfying nonlinear constitutive relationships arise in numerous engineering, chemical and materials processing applications and geophysical situations. Simulations of such flows pose numerical challenges especially in 3D, whose adequate resolution are associated with high computational demands. Lattice Boltzmann methods (LBM) are inherently parallel approaches for flow simulations. Prior efforts in using such kinetic schemes have mainly focused on computing non-Newtonian flows in 2D, and generally utilizing less robust underlying collision models. In this work, we present a new 3D cascaded LBM for the simulation of non-Newtonian power law fluids on a D3Q19 lattice. The effect of collisions are prescribed in terms of changes in different central moments using multiple relaxation times, whose derivation is given in detail. In particular, the relaxation times of the second order moments are related to the local shear rate and parameterized by the model coefficient and exponent representing the power law fluids. We performed a validation of our 3D central moment LBM for various benchmark problems including the complex non-Newtonian flow inside a cubic cavity driven by its top lid at Reynolds numbers of 100, 400 and 1000, and with the power law index specified to be equal to 0.8, 1.0 and 1.5. Comparisons against recent numerical benchmark solutions showed very good agreement. We demonstrated significant improvements in numerical stability allowing simulations of 3D shear driven flows of non-Newtonian fluids (i.e., shear thinning and shear thickening) at higher fluid velocities or Reynolds numbers with the use of the 3D cascaded LBM when compared to other common LB formulations, such as the SRT-LBM and the standard MRT-LBM based on raw moments. The use of central moments relaxations during the collision step naturally enforces the Galilean invariance of the independent moments supported by the lattice with higher order velocity terms in its equilibria which contributes to its superior stability characteristics. The new expressions for the strain rate tensor (Eq.~(\ref{eq:strainratetensor})) based on the non-equilibrium moments for the 3D cascaded LBM presented in this work for prescribing the constitutive relations are local in nature, unlike those for a recent LB scheme for 3D non-Newtonian flows~\cite{chen2020simplified} which are based on finite difference schemes, and hence facilitates parallel computation. Moreover, our approach avoids the various approximations made in Ref.~\cite{chen2020simplified} that can compromise its overall accuracy, is shown to be second order accurate, and has additional degrees of freedom in terms of relaxation parameters to independently adjust the various transport coefficients (i.e., shear and bulk viscosities) for non-Newtonian fluid flows. It may be noted that the 3D cascaded LBM presented in this work (Sec.~\ref{sec:Threedimensionalcascaded}) along with the local expressions for the strain rate tensor (Eq.~(\ref{eq:strainratetensor})) for robust simulation of non-Newtonian flows readily generalizes to other types of essentially 3D flow problems involving spatial variations in effective viscosity such as those involving subgrid scale models dependent on the strain rate tensor.

\section*{Acknowledgements}
The last two authors would like to acknowledge the support of the US National Science Foundation (NSF) under Grant CBET-1705630.


\end{document}